\documentclass[prl,twocolumn,superscriptaddress,floatfix,10pt]{revtex4-1}
\usepackage{amsfonts}
\usepackage{amsmath}
\usepackage{amsthm}
\usepackage{mathtools}
\usepackage{longtable}
\usepackage{amssymb}
\usepackage{amsbsy}
\usepackage{graphicx}
\usepackage{bm}
\usepackage{color}
\usepackage{mathrsfs}
\usepackage[colorlinks,bookmarks=true,citecolor=blue,linkcolor=red,urlcolor=blue]{hyperref}
\usepackage{appendix}
\usepackage{booktabs}
\usepackage[export]{adjustbox}
\usepackage{braket}
\usepackage[makeroom]{cancel}

\usepackage{pdfpages}
\usepackage{pgffor}
\makeatletter
\AtBeginDocument{\let\LS@rot\@undefined}
\makeatother

\begin{document}
\title{Resolution of Topology and Geometry from Momentum-Resolved Spectroscopies}
\author{Shaofeng Huang}
\affiliation{Beijing National Laboratory for Condensed Matter Physics and Institute of Physics, Chinese Academy of Sciences, Beijing 100190, China}
\affiliation{University of Chinese Academy of Sciences, Beijing 100049, China}

\author{Chen Fang}
\email{cfang@iphy.ac.cn}
\affiliation{Beijing National Laboratory for Condensed Matter Physics and Institute of Physics, Chinese Academy of Sciences, Beijing 100190, China}
\affiliation{Kavli Institute for Theoretical Sciences, Chinese Academy of Sciences, Beijing 100190, China}

\begin{abstract}
    Extracting the complete quantum geometric and topological character of Bloch wavefunctions from experiments remains a challenge in condensed matter physics. Here, we resolve this by introducing the ``wavefunction form factor" (WFF) matrix, a quantity directly constructible from intensities in momentum- and energy-resolved spectroscopies like ARPES and INS. We demonstrate that band topology is encoded in ``spectral nodes"---momentum-space points where the WFF determinant vanishes, providing a direct readout of topological invariants via a topological selection rule. Furthermore, when the number of independent probes exceeds the number of the target bands, our framework yields an effective band projector. This enables the determination of Wilson loop spectra and the extraction of an effective quantum geometric tensor, providing a model-independent measurement of the non-Abelian Berry curvature and quantum metric as resolved by the experimental probes.
\end{abstract}

\maketitle

\textit{Introduction.---}
The modern understanding of topological quantum phases in solids is built upon geometric properties of Bloch wavefunctions. This is characterized in two levels: global topology~\cite{hasan2010colloquium,qi2011topological,chiu2016classification} and local geometry~\cite{yu_quantum_2025}. Topology manifests as quantized invariants, such as Chern numbers~\cite{thouless1982quantized} and $Z_2$ indices~\cite{Z2_Fu}, which classify different phases and predict exotic phenomena like quantum Hall effect~\cite{klitzing1980new} and protected boundary modes~\cite{QSHE_Kane}. More recently, focus has been broadened to the underlying local properties of wavefunctions, termed quantum geometry. This is described by the complex quantum geometric tensor (QGT)~\cite{provost_riemannian_1980,yu_quantum_2025}, which contains the Berry curvature (the local source of topology) as its imaginary part and the quantum metric as its real part---a quantity increasingly recognized as crucial for understanding phenomena in flat bands~\cite{peotta_superfluidity_2015,PhysRevLett.128.087002}, fractional Chern insulators~\cite{PhysRevB.93.235133}, and nonlinear optics~\cite{Ahn2022}. Despite this theoretical progress, extracting wavefunction geometry directly from experiment remains a fundamental challenge.

Momentum-resolved spectroscopies---angle-resolved photoemission spectroscopy (ARPES) for electronic bands~\cite{lv_experimental_2015,hasan_weyl_2021,xu_experimental_2015,lv_observation_2015,yao_observation_2019,xu_discovery_2015,lv_observation_2015-1,xu_observation_2016,xu_discovery_2015-1,neupane_observation_2014,liu_stable_2014,borisenko_experimental_2014,xu_observation_2015,liu_discovery_2014,deng_experimental_2016,lu_experimental_2015} and inelastic neutron scattering (INS) for bosonic bands~\cite{zhang_double-weyl_2018,scheie_dirac_2022,yao_topological_2018,yuan_dirac_2020,jin2022chern,miao2018observation,chisnell2015topological}---are premier tools for mapping band structures. 
While they excel at validating band topology by visualizing boundary modes and bulk band crossings, extracting bulk wavefunction topology and geometry is difficult. Recent advances include: (1) \cite{kim2025direct} extracted the quantum metric from polarized ARPES; (2) \cite{Kang_2024} reconstructed QGT via the experimentally accessible quasi‑QGT; (3) \cite{jin2022chern,scheie_dirac_2022} identified a relation between the intensity modulation near the band crossing and the relevant topological invariants.
Notably, these approaches rely on a two-band approximation and only determine the Abelian geometric quantities or topological character within limited regions of the Brillouin zone.

In this Letter, we provide a model-independent framework enabling the direct resolution of band topology and quantum geometry from the spectroscopic intensities. We introduce the ``wavefunction form factor" (WFF) matrix, which is Hermitian and measures the overlap between the band projector $\hat{P}_{\mathbf{k}}$ with a set of experimental probe states. 
The probe states are determined by the experimental configuration (incident particles and scattering momentum transfer) and the static atomic or magnetic structure of the ground state; crucially, probe states are independent of the targeted Bloch states and topologically trivial.
We derive a quantitative method to reconstruct WFF from the spectroscopic interferometry. This connection yields two principal results. First, when the number of the probes ($N_p$) equals the dimension of band projector ($N_b$), topological band crossing with nontrivial charges $\sum_i Q_i$ induces ``spectral nodes"---enforced zeros in the WFF determinant. The spectral nodes come from the topological mismatch between Bloch wavefunctions and the probe states, and we dub this ``topological selection rule."~\cite{huang2025superdiffusive} Moreover, a quantitative relation between the total charge $\sum_{i \in V} Q_i$ and the total degrees of the nodes $\sum_{i \in \partial V} d(\mathbf{k}_i)$ on the surface is derived 
\begin{equation}
    \sum_{i \in V} Q_i = \sum_{i\in \partial V} d(\mathbf{k}_i),
\end{equation}
with equality understood modulo $2$ for $Z_2$ indices. Second, when the number of independent probes exceeds the rank of the band projector, \(N_p > N_b\), the measured WFF matrix yields an effective Bloch projector \(\Pi_{\mathbf{k}}\) that is topologically equivalent to \(\hat{P}_{\mathbf{k}}\) and therefore carries the same topological invariants.
$\Pi_{\mathbf{k}}$ gives an effective QGT, which provides the quantum metric tensor and non-Abelian Berry curvature resolved by the probe states, and with the Berry curvature, we obtain the Wilson-loop spectrum and the associated hybrid-Wannier spectral flow, which faithfully diagnose the band topology.

\begin{figure}[t]
    \includegraphics[width=1.0\linewidth]{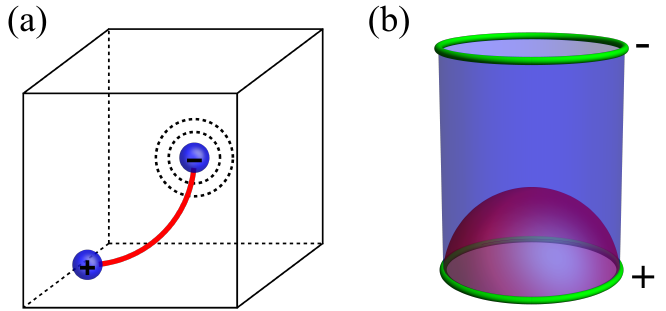}
    \caption{\textbf{Geometry of the nodal manifold enforced by the selection rule}.
    (a) For point charges (e.g. Weyl points) of opposite signs, spectral nodes coalesce into a one-dimensional nodal arc (red) that connects the two monopole charges; along this nodal arc the ARPES/INS intensity vanishes.
    (b) For nodal lines (green loops), the zeros from the $\pi$ Berry phase assemble into a two-dimensional nodal membrane (red) bounded by the line and the zeros from the $Z_2$ monopole charge form a nodal tube (blue) connecting nodal lines with opposite charges.}
    \label{fig:illustration}
\end{figure}

\begin{figure*}[htbp]
    \includegraphics[width=0.95\linewidth]{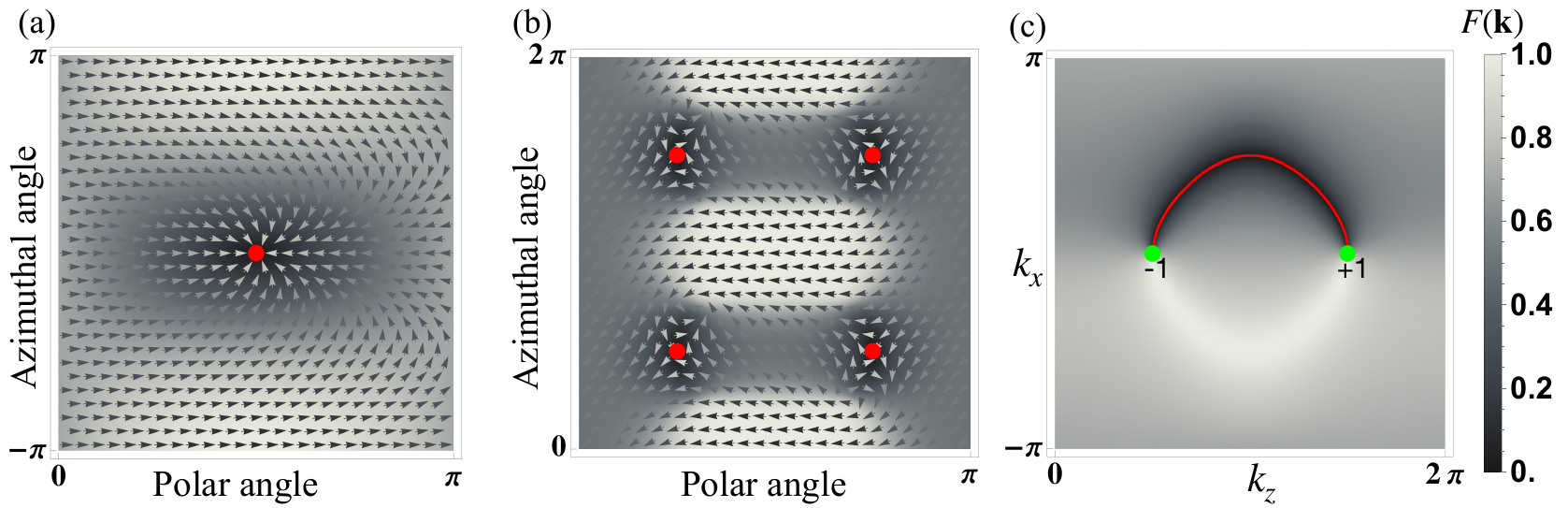}
    \caption{
    \textbf{Topological selection rule and nodal arc for Weyl points.} The wavefunction form factor (WFF) is calculated on different momentum space manifolds. The density plot shows the $F(\mathbf{k})$, while the vector field indicates the phase winding of $\braket{u_{n\mathbf{k}}|W_{a\mathbf{q}}}$. 
    (a) On a sphere enclosing a charge-1 Weyl point, a single spectral node (red circle) appears, around which the phase winds by $2 \pi$.
    (b) On a sphere enclosing a charge-4 quadrupole Weyl point, four order-1 spectral nodes (red circles) appear. The phase winds by $2 \pi$ around each node (winding number $d=1$), satisfying the relation $\sum_i d_i =4 = C.$
    (c) On a 2D slice of Brillouin zone ($\mathbf{k} = (k_x,0,k_z)$) that intersects a pair of Weyl points located at $(0,0, \pi/2 / 3\pi/2)$ with opposite charges ($C = \pm 1$), a line of nodes---the spectral nodal arc (red line)---is visible. The arc connects Weyl points with opposite charges.
    }
    \label{fig:Weyl}
\end{figure*}

{\textit{Wavefunction form factor}.---}
Our framework is built upon the wavefunction form factor (WFF), which we define to capture the wavefunction information encoded in spectroscopic intensities. The target degenerate bands of $N_b$ Bloch states, assumed to be energetically separated from other bands, are described by the projector $\hat{P}_{\mathbf{k}} = \sum_{n = 1}^{N_b} \ket{u_{n\mathbf{k}}} \bra{u_{n\mathbf{k}}}$ where $\ket{u_{n\mathbf{k}}}$ denotes the periodic part of the Bloch wavefunction (wavefunction hereafter). The corresponding WFF is a Hermitian matrix of the overlaps with a set of $N_p$ experimental probe states $\{\ket{W_{a\mathbf{q}}}\}$ that depend on the scattering momentum transfer $\mathbf{q}$:
\begin{equation}
    F_{ab}(\mathbf{q}) = \braket{W_{a\mathbf{q}}|\hat{P}_{\mathbf{k}}|W_{b\mathbf{q}}}.
\end{equation}
Here, the scattering momentum $\mathbf{q}$ and crystal momentum $\mathbf{k}$ are related $\mathbf{q} = \mathbf{k}+\mathbf{G}$ for a reciprocal lattice vector $\mathbf{G}$. In standard spectroscopic geometries, $\mathbf{G}$ is fixed, allowing us to treat the WFF as a function of $\mathbf{k}$ alone~\cite{jin2022chern}.

The physical nature of the probe states $\{\ket{W_{a\mathbf{q}}}\}$ is specific to the measurement and the index $a$ labels the probe states with different experimental configurations. In ARPES, the intensity is governed by the dipole matrix elements $\braket{\chi_{\mathbf{q}\sigma}|\boldsymbol{\epsilon}\cdot \hat{\mathbf{v}}|\psi_{n\mathbf{k}}}$~\cite{sobota2021angle}, where $\ket{\chi_{\mathbf{q}\sigma}}$ is the final photoelectron state with spin $\sigma$, $\boldsymbol{\epsilon}$ is the incident light polarization vector, $\hat{\mathbf{v}}$ is the velocity operator and $\ket{\psi_{n\mathbf{k}}} = e^{i\mathbf{k}\cdot \hat {\mathbf{r}}} \ket{u_{n\mathbf{k}}}$. Consequently, the probe states $\ket{W_{a\mathbf{q}}}$ are identified with $e^{-i \mathbf{k}\cdot \hat{\mathbf{r}}} (\boldsymbol{\epsilon}\cdot \hat{\mathbf{v}})^\dagger \ket{\chi_{\mathbf{q}\sigma}}$ where the probe index $a$ is a combined index of light polarization $\boldsymbol{\epsilon}$ and electron spin $\sigma$. In INS, $\ket{W_{a\mathbf{q}}}$ are determined by the momentum transfer $\mathbf{q}$ and the static system structure, including equilibrium atomic positions, nuclear scattering lengths and the magnetic structure, i.e., magnetic form factor and ground state spin orientations~\cite{lovesey1986theory,squires1996introduction}, see~\cite{suppmat} for details. The probe index $a$ labels the Cartesian spatial components of the magnetic dipole operator perpendicular to $\hat{\mathbf{q}}$.
Provided $\mathbf{G} \neq 0$ the probe states $\{\ket{W_{a\mathbf{q}}}\}$ are topologically trivial in both cases.

The observable quantity---the scalar intensity $\mathcal{I}_c$---measured in a single experimental configuration labeled by $c$ is determined by the probe state (described by a density matrix $\rho^{\text{probe}}_c$), and the ``intensity matrix" $I(\mathbf{k},\omega)$:
\begin{equation}
    \mathcal{I}_c(\mathbf{k},\omega) = \operatorname{Tr}[{\rho}^{\text{probe}}_c I(\mathbf{k},\omega)]. \label{eq:WFF_intensity}
\end{equation}
In general cases, $I_{ab}$ decomposes as $I_{ab}(\mathbf{k},\omega) = \sum_m F_{ab}^{(m)}(\mathbf{k})A_m(\mathbf{k},\omega)$ where the index $m$ denotes different groups of energetically separated bands and $F^{(m)}(\mathbf{k}),A_m(\mathbf{k})$ are the corresponding WFF matrix and spectral function. While the spectral weight may generally mix contributions from multiple bands, our framework focuses on a target group of bands $m_0$. The extraction of $F^{(m_0)}$ is exact in the non-interacting limit (where $A_m$'s are $\delta$-functions) and remains robust for interacting systems provided the target bands dominate the spectral weight at the measured frequency. Under this condition, $I_{ab}(\mathbf{k},\omega) \approx F_{ab}^{(m_0)}(\mathbf{k})A_{m_0}(\mathbf{k},\omega)$; we suppress the index $m_0$ hereafter.

To fully reconstruct the matrix structure of the intensity matrix from scalar intensities, we exploit the fact that the set of experimentally realizable probe density matrices spans the space of $N_p \times N_p$ Hermitian matrices. We adopt a decomposition scheme using a complete orthogonal basis of Hermitian matrices $\{\Gamma_{\mu}\}$ (including the identity).
Specifically: for INS, the probe space is 2D ($N_p=2$) and $\{\Gamma_\mu\}$ are the Pauli matrices; for spin- and polarization-resolved ARPES, the probe space is 4D ($N_p=4$) and $\{\Gamma_\mu\}$ corresponds to the tensor products of spin and orbital pseudo-spin matrices, $\Gamma_\mu = \sigma_i \otimes \tau_j$ with the composite index $\mu = (i,j)$. We normalize this basis such that $\operatorname{Tr}[\Gamma_\mu \Gamma_{\nu}] = N_p \delta_{\mu \nu}$. The WFF matrix is expanded as $F(\mathbf{k}) = \sum_{\mu} f_{\mu}(\mathbf{k}) \Gamma_{\mu}$. 
Since the set of experimental probe configurations $\{\rho^{\text{probe}}_c\}$ is complete, we relate the probe density matrices to the Pauli basis via an invertible transformation matrix $B$, such that $\rho^{\text{probe}}_c = \sum_{\mu} B_{c \mu} \Gamma_{\mu}$.
Substituting this into the trace relation yields the linear system $\mathcal{I}_c(\mathbf{k},\omega) = N_p A(\mathbf{k},\omega) \sum_{\mu} B_{c\mu}f_{\mu}(\mathbf{k})$. 
Consequently, the spectral-weighted coefficients are extracted via matrix inversion: $f_{\mu}(\mathbf{k}) A(\mathbf{k},\omega)  = \frac{1}{ N_p} \sum_c [B^{-1}]_{\mu c}\mathcal{I}_c(\mathbf{k},\omega)$. 
Physically, this demonstrates that each component of the WFF (weighted by the spectral function) is determined by a specific linear combination of measured intensities. 

Physically, this reconstruction is achieved by explicitly controlling the degrees of freedom of the incident and scattered particles. In spin- and polarization-resolved ARPES, the probe density matrix factorizes as ${\rho}^\mathrm{probe} = \rho^{\sigma} \otimes \rho^{\tau}$~\cite{sobota2021angle}. Here, $\rho^{\sigma} =\frac{1}{2}(1 +  \hat{\mathbf{P}}_e \cdot \boldsymbol{\sigma})$ describes the spin-resolving detector configuration with polarization axis $\hat{\mathbf{P}}_e$, while $\rho^{\tau} = \frac{1}{2}(1 + \mathbf{S} \cdot \boldsymbol{\tau})$ describes the incident light polarization, parameterized by the Stokes vector $\mathbf{S}$ ($|\mathbf{S}| \le 1$)~\cite{Gil_2007}. 
Because the detector spin axis $\hat{\mathbf{P}}_e$ and the photon Stokes vector $\mathbf{S}$ can be varied independently, this tensor product structure naturally spans the full space of $4 \times 4$ Hermitian matrices required to construct the $\sigma_i \otimes \tau_j$ basis. 
Similarly, in INS, the probe density matrix is explicitly given by $\rho^{\text{probe}} = \frac{1}{2}[C_0(\mathbf{P}_{\text{in}},\hat{\mathbf{q}}) + \hat{\mathbf{{P}}}_{\text{out}} \cdot \mathbf{C}(\mathbf{P}_{\text{in}},\hat{\mathbf{q}})]$. Here, $C_0$ and $\mathbf{C} = (C_x, C_y, C_z)$ constitute a set of four Hermitian kinematic matrices~\cite{suppmat} determined by the incident neutron polarization $\mathbf{P}_{\text{in}}$ and the momentum transfer direction $\hat{\mathbf{q}}$, and $\hat{\mathbf{P}}_{\text{out}}$ is the polarization analyzer axis. These  matrices form a complete basis for the $2\times2$ Hermitian matrices. 
Consequently, by tuning the polarization analyzer axis $\hat{\mathbf{P}}_{\text{out}}$, the resulting probe density matrix sweeps out the full $2 \times 2$ Hermitian space required to reconstruct the WFF~\cite{suppmat}.
In both techniques, systematically varying these control parameters generates the complete set of linearly independent intensity measurements required to invert the $B$ matrix and determine $F(\mathbf{k})$.

{\textit{Topological Selection Rule}---}
We now demonstrate that the WFF provides a direct measurement of the band topology when the number of probes equals the number of target bands ($N_p = N_b$). A nontrivial topological invariant obstructs the existence of a globally smooth gauge for wavefunctions. This creates a topological mismatch with the trivial experimental probe states, enforcing their overlap or the WFF determinant, $\det[F(\mathbf{k})]$, to vanish at specific points on the closed surface enclosing the topological band crossing. These enforced zeros, or ``spectral nodes," are direct fingerprints of the underlying topology: this is the topological selection rule~\cite{huang2025superdiffusive}.

Consider first a single band $(N_b = 1)$ hosting a Weyl point with the Chern number $C$.
Now the WFF, $F_{aa}(\mathbf{k}) = |\braket{W_{a\mathbf{q}}|u_{n\mathbf{k}}}|^2$, is a scalar. The nonzero Chern number is an obstruction to finding a globally smooth gauge for the wavefunction $\ket{u_{n\mathbf{k}}}$ on the sphere enclosing the Weyl point~\cite{bernevig2013topological}. In contrast, the probe state $\ket{W_{a\mathbf{q}}}$ is topologically trivial. This mismatch leads to zeros of $F_{aa}(\mathbf{k})$ on the enclosing surface (and consequently the measured intensity) and we prove this by contradiction. Suppose $F_{aa}(\mathbf{k})$ is globally nonzero on the surface, then we can do a gauge transformation
\begin{equation}
    \ket{u_{n\mathbf{k}}}' = \frac{\braket{u_{n\mathbf{k}}|W_{a\mathbf{q}}}}{|\braket{u_{n\mathbf{k}}|W_{a\mathbf{q}}}|} \ket{u_{n\mathbf{k}}} = \frac {\hat{P}_{\mathbf{k}} \ket{W_{a\mathbf{q}}}}{[\braket{W_{a\mathbf{q}}|\hat{P}_{\mathbf{k}}|W_{a\mathbf{q}}}]^{1/2}}. 
\end{equation}
By definition, $\hat{P}_{\mathbf{k}}$ is globally smooth on the enclosing surface, so is $\ket{u_{n\mathbf{k}}}'$, contradicting the non-zero Chern number. More quantitatively~\cite{suppmat}, the Chern number is precisely the sum of the integer winding numbers (degrees) $d(\mathbf{k}_i)$ of the spectral nodes $\mathbf{k}_i$:
\begin{equation}
    C = \sum_{i} d(\mathbf{k}_i),  \label{eq:Chern_sum_rule}
\end{equation}
where $d(\mathbf{k}_i) = \frac{1}{2\pi i} \oint d \log \braket{u_{n\mathbf{k}}|W_{a\mathbf{q}}}$.
This direct correspondence is visualized in Fig.~\ref{fig:Weyl}(a,b), where we demonstrate the WFF for Weyl points of charge $C =1$ and $C=4$ and visualizing the phase winding. In a lattice system, Weyl points always come as pairs with opposite charges. As the enclosing surface is deformed, these nodes trace out ``spectral nodal arcs" connecting those Weyl points with opposite charges.  In Fig.~\ref{fig:Weyl}(c), we illustrate the spectral nodal arc for a lattice system with charge-$1$ Weyl point. Moreover, the discussions about the Weyl points formed by non-degenerate bands can be easily generalized to nodal lines protected by $\pi$-Berry phase, where the zeros of the WFF form a spectral nodal membrane bounded by the nodal line, see Fig.~\ref{fig:illustration}(b). 

This logic can be extended to multi-band systems.
We consider a pair of bands $(N_b=2)$ in a system with composite inversion ($P$) and time-reversal symmetry ($T$) satisfying $(PT)^2 = 1$. In this class one can choose a ``real" gauge where the $PT$-operation acting on the orbitals is simply a complex conjugation and the wavefunctions $\{\ket{u_{n\mathbf{k}}}\}$ are real.
The generic band crossing is a nodal line~\cite{fang2016topological}, which is characterized by two independent $Z_2$ indices---the Berry phase on the linking loop and the $Z_2$ monopole charge on the enclosing sphere (the second Stiefel-Whitney class)~\cite{fang2015topological,ahn2018band}. The $Z_2$ monopole charge is nontrivial only in multi-band systems and is created/annihilated in pairs. 

Similar to the Chern number, the $Z_2$ monopole charge acts as an obstruction to finding a globally smooth, real gauge for the wavefunctions over the sphere enclosing the charge~\cite{fang2015topological}. 
Suppose the probe states $\{\ket{W_{a\mathbf{q}}}\}$ are compatible with the $PT$-symmetry and thus can be chosen to be globally smooth and real-valued.
The topological obstruction due to the $Z_2$ charge manifests as enforced zeros of $\det[F(\mathbf{k})]$, otherwise we can use the WFF to find a globally smooth real gauge for the wavefunctions, contradicting the nontrivial monopole charge~\cite{suppmat}.
Unlike the Weyl case, the presence of locally real gauge implies that these spectral nodes generically form one-dimensional ``nodal loop" on the enclosing sphere. As the sphere is deformed, these loops sweep out a ``nodal tube" connecting nodal lines of opposite $Z_2$ charge in momentum space.

\begin{figure}[t]
    \includegraphics[width=1.0\linewidth]{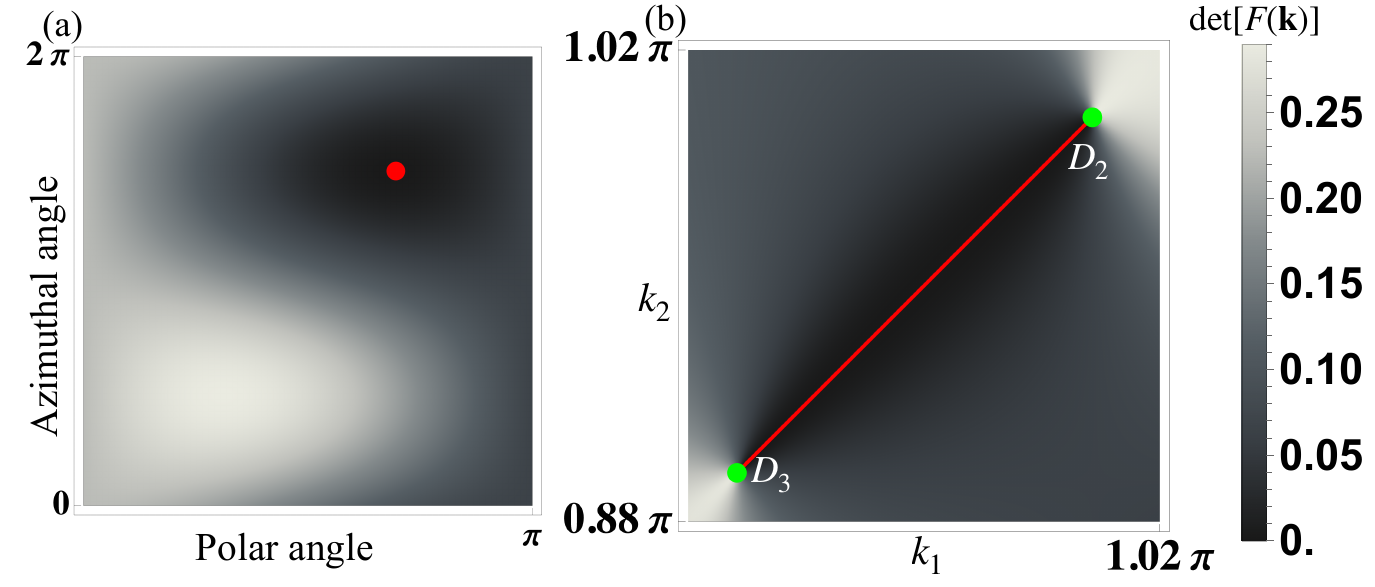}
    \caption{\textbf{Spectroscopic signature of the $Z_2$ charge in $\text{Cu}_3\text{TeO}_6$} The calculated magnitude of the WFF determinant, $\det[F(\mathbf{k})]$, reveals the spectroscopic signatures of the $Z_2$ charge. 
    (a) On a sphere enclosing the Dirac point $D_2 = (\pi,\pi,\pi)$, a spectral node is visible (red circle), confirming the selection rule for a non-trivial $Z_2$ charge. 
    (b) The $\det[F(\mathbf{k})]$ calculated on the 2D momentum space slice defined by $\mathbf{k}=(k_1,k_1,k_2)$. A nodal arc (red line) connects the Dirac point $D_2$ with its partner $D_3$ (labeled by magenta circles).
    In the simulation, we choose the reciprocal lattice vector $\mathbf{G} = 2\pi\times(2,2,2)$, which lies in the [111] rotation axis which is compatible with the system's rotation symmetry. And we see the nodal arc entirely lies in the rotation axis.
    }
    \label{fig:magnon}
\end{figure}

This $Z_2$ monopole charge can arise in (i) electronic band systems without spin-orbit coupling~\cite{fang2015topological}; (ii) phonon band systems~\cite{li2025general}, and (iii) antiferromagnetic magnon band systems~\cite{li2017dirac}.
Realizing the two-probe protocol is challenging for phonons~\cite{suppmat}, but is naturally suited to spin-orbit-coupling free electronic bands (via p/s-type light polarization in ARPES) and magnon bands (via neutron spin in INS). 
And we exemplify the extraction of the $Z_2$ monopole charge in the antiferromagnet $\text{Cu}_3\text{TeO}_6$ which hosts Dirac/nodal line magnon with nontrivial $Z_2$ charge~\cite{li2017dirac}. 
By the full polarization analysis, the WFF determinant can be reconstructed from the standard four-component intensity vector $\mathcal{I}(\mathbf{q},\mathbf{P},\omega) = (\mathcal I_0, {\mathcal I}_x,{\mathcal I}_y,{\mathcal I}_z)^T $, where $\mathbf{P}$ denotes the polarization of the incident neutrons. Here, $\mathcal{I}_0$ is the total cross section while the spatial components are defined by the intensity difference $\mathcal{I}_{j} = \mathcal{I}_{j,+} - \mathcal{I}_{j,-}$, where $\mathcal{I}_{j,\pm}$ denotes the intensity measured with the polarization analyzer aligned parallel ($+$) or anti-parallel ($-$) to the $j$-axis.
For the fully polarized incident neutrons $|\mathbf{P}| = 1$, on the equal energy surface the determinant of the intensity matrix is derived as:
\begin{equation}
    \det[I(\mathbf{k})] = \frac{\mathcal{I}_0^2 - \sum_{j=x,y,z} \mathcal{I}_j^2 }{1- (\hat{\mathbf{q}}\cdot\mathbf{P})^2} ,
\end{equation}
valid away from the geometric singularity $\hat{\mathbf{q}} \parallel \mathbf{P}$.
Crucially, because the intensity matrix relates to the WFF via a scalar spectral function, $I(\mathbf{k},\omega) = F(\mathbf{k})A(\mathbf{k},\omega)$, their determinants share the same nodal structure on the quasiparticle band $(A>0)$.
Thus, the zeros of the experimentally derived $\det[I]$ pinpoint the topological spectral nodes of $\det[F]$.
Then we simulate the $\det[F(\mathbf{k})]$ in $\text{Cu}_3 \text{TeO}_6$ in Fig.~\ref{fig:magnon}. Note that in this material, the approximate $U(1)$ spin-rotation symmetry ($S_z$ conservation) pinches the generic nodal lines into Dirac points. Consequently, the spectral nodal loops predicted for the generic $Z_2$ case also contract into discrete spectral nodes. Fig.~\ref{fig:magnon}(a) shows $\det[F(\mathbf{k})]$ on the sphere enclosing the Dirac point, which exhibits the predicted spectral nodal point. Also on a 2D slice $\mathbf{k} = (k_1,k_1,k_2)$ of the BZ, these nodes trace a continuous nodal arc, connecting two Dirac points of opposite $Z_2$ charge.

\textit{Extraction of the Wilson loop and quantum geometry. ---}
When the number of independent probes exceeds the number of bands ($N_p > N_b$), the redundancy in the spectroscopic data enables us to go beyond the simple detection of spectral nodes. The WFF allows us to construct an effective band projector in the probe space, from which we can extract the effective non-Abelian quantum geometry and the Wilson loop spectra.

We begin by defining the projected wavefunction components $[\phi_{n\mathbf{k}}]_a = \braket{W_{a\mathbf{q}}|u_{n\mathbf{k}}}$. The WFF is then given by $F(\mathbf{k}) = \sum_n \phi_{n\mathbf{k}} \phi^\dagger_{n\mathbf{k}}$. The eigenvectors of $F(\mathbf{k})$ encode the orientation of the wavefunctions within the Hilbert space spanned by the probe states. To isolate this geometric information, we ``flatten'' the WFF by normalizing the $N_b$ non-zero eigenvalues to unity, yielding a rank-$N_b$ projector:
\begin{equation}
    \Pi_{\mathbf{k}}=\sum_{m=1}^{N_b}\tilde{u}_{m\mathbf{k}} \tilde{u}_{m\mathbf{k}}^\dagger,
\end{equation}
where $\{\tilde{u}_{m\mathbf{k}}\}$ are the eigenvectors of $F(\mathbf{k})$ associated with the non-zero eigenvalues. 
This flattening procedure renders the effective projector independent of the spectral weight: because the intensity matrix $I$ and the WFF $F$ share the same eigenvector (differing only by the scalar spectral weight $A$ in their eigenvalues), $\Pi_{\mathbf{k}}$ can be directly read from the intensity matrix without knowing the explicit spectral function.
Provided the no-rank-drop condition holds (i.e., \(\mathrm{rank}\,F(\mathbf{k}) = N_b, \forall \mathbf{k}\)), the map \(\hat P_{\mathbf{k}}\mapsto \Pi_{\mathbf{k}}\) constitutes a vector-bundle isomorphism~\cite{suppmat}. Physically, this implies that $\Pi_{\mathbf{k}}$ is a smooth deformation of the original band projector $\hat{P}_{\mathbf{k}}$~\cite{nakahara2018geometry,milnor1974characteristic} and, consequently, carries identical topological invariants. Note that for symmetry-protected topological phases, this equivalence assumes the probe subspace is compatible with the symmetries, ensuring the deformation remains within the corresponding symmetry class.

From this effective projector, we define the effective non-Abelian quantum geometric tensor (QGT):
\begin{equation}
    \tilde{Q}^{ij}_{mm'} = \tilde{u}_{m\mathbf{k}}^\dagger \partial_i \Pi_{\mathbf{k}} \partial_j \Pi_{\mathbf{k}} \tilde{u}_{m'\mathbf{k}}.
\end{equation}
The Hermitian and anti-Hermitian parts of $\tilde{Q}$ correspond to the effective quantum metric and the effective non-Abelian Berry curvature~\footnote{Note that the spectrum of the effective QGT is fully determined by the effective projector}, respectively~\cite{Ma2010prb}. It is crucial to recognize that $\Pi_{\mathbf{k}}$ represents the wavefunction structure projected onto the subspace spanned by the experimental probe states. Therefore, $\tilde{Q}$ captures the geometry of the wavefunctions \textit{as resolved by the probes}---referred to hereafter as the probe-resolved QGT. Since information orthogonal to the probe subspace is lost during projection, the local geometry is generally distorted relative to the bulk. Indeed, the Berry curvatures corresponding to $\hat{P}_{\mathbf{k}}$ and $\Pi_{\mathbf{k}}$ may differ locally, yet their integrals over the momentum space (the Chern numbers) remain identical~\cite{suppmat}. However, as the number of independent probes $N_p$ increases, the probe subspace expands, and $\tilde{Q}$ eventually converges toward the true bulk geometry~\cite{guillot2025measuring}.

Leveraging the topological equivalence between $\Pi_{\mathbf{k}}$ and $\hat{P}_{\mathbf{k}}$, we construct the Wilson loop operator along a primitive reciprocal vector \(\mathbf{g}\) at fixed \(\mathbf{k}_\perp\):
\begin{equation}
    W_{\mathbf{g}}(\mathbf{k}_\perp)=\lim_{N\to\infty}
    \Big[\,\Pi_{\mathbf{k}_0} \Pi_{\mathbf{k}_1} \cdots \Pi_{\mathbf{k}_{N-1}} \Pi_{\mathbf{k}_\perp + \mathbf{g}}\,\Big],
\end{equation}
with \(\mathbf{k}_j=\mathbf{k}_\perp+j\,\mathbf{g}/N\). The eigenphases of \(W_{\mathbf{g}}(\mathbf{k}_\perp)\) yield the hybrid-Wannier charge centers~\cite{yu2011equiv}. Since the deformation of the projector preserves the topology, this spectral flow provides a robust diagnosis of the bulk topology---such as $Z_2$ indices---despite the local geometric distortions inherent in the effective projector. Examples are provided in~\cite{suppmat}.

\textit{Conclusions.}---
We have introduced the wavefunction form factor (WFF) matrix, a gauge-invariant observable reconstructible from interferometric spectroscopic intensities, to access the topology and geometry of Bloch wavefunctions directly from energy- and momentum-resolved spectroscopies. When the number of probes matches the projector rank ($N_p=N_b$), we established a \emph{topological selection rule}: topological charges enforce zeros (spectral nodes) in the WFF determinant on enclosing surfaces. We proved that the sum of the node degrees equals the total topological charge (modulo 2 for $Z_2$ indices).
These nodes assemble into nodal manifolds---spectral nodal arcs, tubes, and membranes---which serve as direct fingerprints of the underlying topological invariants.
For $N_p>N_b$, flattening the measured WFF yields an effective projector $\Pi_{\mathbf{k}}$ that is topologically equivalent to the band projector $\hat P_{\mathbf{k}}$. This allows the construction of an effective non-Abelian quantum geometric tensor, quantifying the local geometry as resolved by the experimental probes. Crucially, this enables the extraction of robust topological observables, including Wilson loop spectra and hybrid–Wannier center flows.
These results provide a rigorous, model-independent framework to quantify both band topology and quantum geometry directly from experimental intensity data.

\bibliography{main}
\newpage


\section*{Supplementary material}

# Supplemental material for "Resolution of Topology and Geometry from Momentum-Resolved Spectroscopies"

Shaofeng Huang[1,2] and Chen Fang[1,3,*]

[1]*Beijing National Laboratory for Condensed Matter Physics and Institute of Physics,
Chinese Academy of Sciences, Beijing 100190, China*
[2]*University of Chinese Academy of Sciences, Beijing 100049, China*
[3]*Kavli Institute for Theoretical Sciences, Chinese Academy of Sciences, Beijing 100190, China*

## CONTENTS

## I. ARPES FOR ELECTRONIC BAND

### 1. Derivation of the ARPES intensity

In this section, we derive the ARPES intensity from the ab-initio Fermi's Golden rule [1]. We adapt the *three-step* model: (1) The photon drives a direct optical transition for an electron in the bulk. In this step all the momenta

are conserved. (2) The excited electron propgates to the surface. (3) The photoelectron is transmitted through the surface barriers, with the electron ultimately occupying a free-electron final state in the lab vacuum. We also adapt the *sudden approximation*. In this limit, photoelectron has no time to interact with the rest $(N-1)-$electron system once being excited by the photon.

The transition rate given by the Fermi golden rule is

$$\sum_f |w_{fi}|^2 = \frac{2\pi}{\hbar} \sum_f |\langle \Psi_f^N | \hat{H}_{\text{int}} | \Psi_i^N \rangle|^2 \delta(E_i^N + h\nu - E_f^N). \tag{1}$$

By the sudden approximation, the final $N$-electron state can be approximated by a product state of an outgoing final photoelectron state $|\chi_{\mathbf{k}}\rangle$ with kinetic energy $\epsilon_f$ (not to be confused with the Fermi energy) and an $(N-1)-$electron eigenstate $|\Psi_m^{N-1}\rangle$:

$$|\Psi_f^N\rangle \approx |\chi_{\mathbf{q}}\rangle \otimes |\Psi_m^{N-1}\rangle. \tag{2}$$

Within the sudden approximation, the outgoing electron is distinguishable from the $(N-1)-$electron state, so we do not need to anti-symmetrize the product state. The scattering rate can be rewritten as

$$w(\mathbf{q},\nu) = \frac{2\pi}{\hbar} \sum_m |\langle \chi_{\mathbf{q}}; \Psi_m^{N-1} | \hat{H}_{\text{int}} | \Psi_i^N \rangle|^2 \delta(E_i^N + h\nu - E_m^{N-1} - \epsilon_f). \tag{3}$$

Now let's deal with the matrix element. In the Coulomb gauge and within the dipole approximation (the wavelength of light is much larger than the micro-lattice spacing), we have

$$\hat{H}_{\text{int}} = \frac{e}{mc} \mathbf{A} \cdot \hat{\mathbf{p}} = \frac{eA}{mc} \boldsymbol{\epsilon} \cdot \hat{\mathbf{p}}, \tag{4}$$

where $\hat{\mathbf{p}}$ is the momentum operator and $\boldsymbol{\epsilon}$ is the polarization of light. In the presence of SoC, we have

$$\hat{H}_{\text{int}} = \frac{e}{mc} \mathbf{A} \cdot \hat{\mathbf{p}} + \frac{e\hbar}{4m^2c^2} \mathbf{A} \cdot (\boldsymbol{\sigma} \times \nabla V(\mathbf{r})). \tag{5}$$

Without losing generality, we ignore the additional SoC term for simplicity. To write the interaction operator in the second quantized language, we expand the electron field operator

$$\hat{\Psi}(\mathbf{r}) = \sum_\alpha \phi_\alpha(\mathbf{r}) \hat{c}_\alpha + \sum_{\mathbf{q}} \chi_{\mathbf{q}}(\mathbf{r}) \hat{a}_{\mathbf{q}}. \tag{6}$$

Here we divide Hilbert space into two parts: the continuum in the lab ($\chi_{\mathbf{q}}(\mathbf{r})$) and Wannier basis in the crystal ($\phi_\alpha(\mathbf{r})$). And the second quantized interaction operator is given by

$$\hat{H}_{\text{int}} = \int d\mathbf{r}\, \hat{\Psi}^\dagger(\mathbf{r}) \left( \frac{eA}{mc} \boldsymbol{\epsilon} \cdot \hat{\mathbf{p}} \right) \hat{\Psi}(\mathbf{r}). \tag{7}$$

The expanded Hamiltonian contains three parts: crystal to crystal; continuum to continuum; hybridization between continuum and crystal. Below we see only the continuum to crystal part contributes to the scattering rate. The initial state can be rewritten as

$$|\Psi_i^N\rangle \equiv |0\rangle \otimes |\Psi_i^N\rangle, \tag{8}$$

and the matrix element

$$\begin{aligned}
&(\langle \chi_{\mathbf{q}}| \otimes \langle \Psi_m^{N-1}|) \hat{H}_{\text{int}} (|\Psi_i^N\rangle \otimes |0\rangle) \\
&= (\langle \chi_{\mathbf{q}}| \otimes \langle \Psi_m^{N-1}|) \left( \sum_{\mathbf{q}'\alpha} \langle \chi_{\mathbf{q}'}|H_{\text{int}}|\phi_\alpha\rangle \hat{a}_{\mathbf{q}'}^\dagger \hat{c}_\alpha \right) (|\Psi_i^N\rangle |0\rangle) \\
&= \sum_\alpha \langle \Psi_m^{N-1}|\hat{c}_\alpha|\Psi_i^N\rangle \langle \chi_{\mathbf{q}}| \left( \frac{eA}{mc} \boldsymbol{\epsilon} \cdot \hat{\mathbf{p}} \right) |\phi_\alpha\rangle \\
&= \sum_\alpha \langle \Psi_m^{N-1}|\hat{c}_\alpha|\Psi_i^N\rangle M(\mathbf{q}, \alpha, \boldsymbol{\epsilon}).
\end{aligned} \tag{9}$$

Now we sum over all these matrix elements and get

$$\mathcal{I}(\mathbf{q},\omega) = \sum_{\alpha\beta} M(\mathbf{q},\alpha,\boldsymbol{\epsilon})M^*(\mathbf{q},\beta,\boldsymbol{\epsilon}) \sum_m \langle\Psi_0^N|\hat{c}_\beta^\dagger|\Psi_m^{N-1}\rangle \langle\Psi_m^{N-1}|\hat{c}_\alpha|\Psi_0^N\rangle \delta(\omega - (E_i^N - E_m^{N-1})). \quad (10)$$

Now we change to the band basis $\alpha \to (n, \mathbf{k})$. Notice that we use $\mathbf{k}$ to denote the crystal momentum and $\mathbf{q}$ to denote the free momentum. And the measured intensity can be expressed by

$$\mathcal{I}(\mathbf{q},\omega) = \sum_m \sum_{n,\mathbf{k}} \sum_{\mathbf{G}} (M_n(\mathbf{q},\mathbf{k},\boldsymbol{\epsilon})M_n^*(\mathbf{q},\mathbf{k},\boldsymbol{\epsilon})) \langle\Psi_0^N|\hat{\psi}_{n\mathbf{k}}^\dagger|\Psi_m^{N-1}\rangle \langle\Psi_m^{N-1}|\hat{\psi}_{n\mathbf{k}}|\Psi_0^N\rangle \delta(\mathbf{q}-\mathbf{k}-\mathbf{G})\delta(\omega-(E_i^N - E_m^{N-1})). \quad (11)$$

The one particle removal Green's function, by definition, is [2]

$$A_n^{(-)}(\mathbf{k},\omega) = 2\pi \sum_m \langle\Psi_0^N|\hat{c}_{n\mathbf{k}}^\dagger|\Psi_m^{N-1}\rangle \langle\Psi_m^{N-1}|\hat{c}_{n\mathbf{k}}|\Psi_0^N\rangle \delta(\omega - (E_i^N - E_m^{N-1})). \quad (12)$$

And it is given by the lesser Green's function in the frequency domain

$$G^<(\alpha t; \beta t') = i \langle \hat{c}_\beta^\dagger(t')\hat{c}_\alpha(t)\rangle, \quad (13)$$

where $\langle\ldots\rangle$ stands for thermal average. And its relation with the full spectral function (which is given by the imaginary part of the retarded Green's function) is

$$-iG^<(n\mathbf{k},\omega) = A_n(\mathbf{k},\omega)n_F(\omega). \quad (14)$$

Below we will drop the delta-function from the crystal momentum conservation and one should keep in mind $\mathbf{q} = \mathbf{k}$ mod $\mathbf{G}$.

The final intensity now can be factorized into three parts

$$\mathcal{I}(\mathbf{q},\omega) \propto \sum_n M_n(\mathbf{q},\boldsymbol{\epsilon})A_n(\mathbf{k},\omega)M_n^*(\mathbf{q},\boldsymbol{\epsilon})n_F(\omega), \quad (15)$$

where

$$M_n(\mathbf{q},\boldsymbol{\epsilon}) = \langle\chi_\mathbf{q}|\left(\frac{eA}{mc}\boldsymbol{\epsilon}\cdot\hat{\mathbf{p}}\right)|\psi_{n\mathbf{k}}\rangle. \quad (16)$$

We immediately see that the matrix element contains the information we wanted. Now we define three probe states

$$|W_{\alpha\mathbf{q}}\rangle = \frac{eA}{mc}e^{-i\mathbf{k}\cdot\hat{\mathbf{r}}}\hat{p}^\alpha|\chi_\mathbf{q}\rangle, \quad (17)$$

and notice that the Bloch states can be rewritten as $|\psi_{n\mathbf{k}}\rangle = e^{i\mathbf{k}\cdot\hat{\mathbf{r}}}|u_{n\mathbf{k}}\rangle$ where $|u_{n\mathbf{k}}\rangle$ is the periodic part of the Bloch states (wavefunction hereafter), and we denote its projector as $\hat{P}_{n\mathbf{k}} = |u_{n\mathbf{k}}\rangle\langle u_{n\mathbf{k}}|$. Notice that one can also shift to length gauge of the dipole term, where the matrix elements take the form:

$$M_n(\mathbf{q},\boldsymbol{\epsilon}) \propto \langle\chi_\mathbf{q}|\boldsymbol{\epsilon}\cdot\hat{\mathbf{r}}|\psi_{n\mathbf{k}}\rangle. \quad (18)$$

The intensity can be rewritten as

$$\mathcal{I}(\mathbf{q},\omega) = \sum_n \sum_{\alpha\beta} \epsilon^\alpha \epsilon^{\beta*} \langle W_{\alpha\mathbf{q}}|\hat{P}_{n\mathbf{k}}|W_{\beta\mathbf{q}}\rangle A_n(\mathbf{k},\omega)n_F(\omega). \quad (19)$$

Since the light is always transversely polarized, the polarization vector is always perpendicular to the momentum of the photon. Denoting the photon momentum as $\mathbf{Q}$ which is fixed in the photoemission process, we define a local frame. The vertical direction is given by

$$\mathbf{e}_v = \frac{\mathbf{e}_z - \hat{\mathbf{Q}}(\mathbf{e}_z \cdot \hat{\mathbf{Q}})}{|\mathbf{e}_z - \hat{\mathbf{Q}}(\mathbf{e}_z \cdot \hat{\mathbf{Q}})|}, \quad (20)$$

and $\mathbf{e}_h = \hat{\mathbf{Q}} \times \mathbf{e}_v$. And we can define four independent light polarization: (1) Linear Horizontal (LH) $\boldsymbol{\epsilon}_{LH} = \mathbf{e}_h$; (2) Linear Vertical (LV) $\boldsymbol{\epsilon}_{LV} = \mathbf{e}_v$; (3) Left (Right)-hand Circular polarized (LC/RC) $\boldsymbol{\epsilon}_{L/RC} = \frac{1}{\sqrt{2}}(\mathbf{e}_h \pm i\mathbf{e}_v)$; (4) Diagonal (anti-)polarized (DP/ADP) $\boldsymbol{\epsilon}_{DP} = \frac{1}{\sqrt{2}}(\mathbf{e}_h + \mathbf{e}_v)$.

Furthermore, the spin-resolved ARPES intensity is given by post-selecting the spin of the outgoing photoelectron and the spin-resolved ARPES intensity is given by

$$\begin{aligned}
\mathcal{I}(\mathbf{q},\omega) &= \frac{1}{2}\sum_n \sum_{\alpha\beta}\sum_{\sigma,\sigma'} \epsilon^\alpha \epsilon^{\beta*}(1\pm\boldsymbol{\sigma})_{\sigma\sigma'} \langle W_{(\alpha,\sigma')\mathbf{q}}|\hat{P}_{n\mathbf{k}}|W_{(\beta,\sigma)\mathbf{q}}\rangle A_n(\mathbf{k},\omega)n_F(\omega) \\
&= \sum_n \sum_{\alpha\beta}\sum_{\sigma,\sigma'} (\rho^{\boldsymbol{\tau}}\otimes\rho^{\boldsymbol{\sigma}})_{\beta\sigma,\alpha\sigma'} \langle W_{(\alpha,\sigma')\mathbf{q}}|\hat{P}_{n\mathbf{k}}|W_{(\beta,\sigma)\mathbf{q}}\rangle A_n(\mathbf{k},\omega)n_F(\omega) \\
&= \sum_n \text{Tr}[\rho^{\boldsymbol{\tau}}\otimes\rho^{\boldsymbol{\sigma}} F^{(n)}(\mathbf{k})] A_n(\mathbf{k})n_F(\omega)
\end{aligned} \quad (21)$$

where the density matrix of the joint light-spin polarization is given by

$$\rho^{\boldsymbol{\tau}} \otimes \rho^{\boldsymbol{\sigma}} = (\frac{1}{2})^2 (1\pm\boldsymbol{\tau})\otimes(1\pm\boldsymbol{\sigma}). \quad (22)$$

Here, if $\rho^{\boldsymbol{\tau}} = \frac{1}{2}(1+\tau^x)$, it means the incident light is linearly diagonal polarized (DP), instead of being polarized along the $+x$-direction. Similarly, if $\rho^{\boldsymbol{\epsilon}} = \frac{1}{2}(1+\tau^y)$, it means the incident light is right-hand circular polarized (LC).

## 2. Extraction of projector's matrix element

On the equal energy surface, $\omega = \varepsilon_{n\mathbf{k}}$, the ARPES intensity is proportional to the matrix elements

$$\mathcal{I}(\mathbf{q},\omega) \propto \text{Tr}[(\rho^{\boldsymbol{\tau}}\otimes\rho^{\boldsymbol{\sigma}})F^{(n)}(\mathbf{k})], \quad (23)$$

In this subsection we show how to extract all the matrix elements $[F^{(n)}]_{\alpha\sigma,\beta\sigma'}$ by performing ARPES measurements with different light polarization and photoelectron spin.

We choose the basis probe state in the photon part as: $\{\rho^{\boldsymbol{\tau}}_{DP}, \rho^{\boldsymbol{\tau}}_{RC}, \rho^{\boldsymbol{\tau}}_{LH}, \rho^{\boldsymbol{\tau}}_{LV}\}$. And the basis states in the spin part are $\{\rho^{\boldsymbol{\sigma}}_{+x}, \rho^{\boldsymbol{\sigma}}_{+y}, \rho^{\boldsymbol{\sigma}}_{+z}, \rho^{\boldsymbol{\sigma}}_{-z}\}$. Then we decompose these matrices in terms the Pauli matrices: $\tau_i \otimes \sigma_j$:

$$\begin{pmatrix} \rho^{\boldsymbol{\tau}}_{DP} \\ \rho^{\boldsymbol{\tau}}_{RC} \\ \rho^{\boldsymbol{\tau}}_{LH} \\ \rho^{\boldsymbol{\tau}}_{LV} \end{pmatrix} = \frac{1}{2}\begin{pmatrix} 1 & 1 & 0 & 0 \\ 1 & 0 & 1 & 0 \\ 1 & 0 & 0 & 1 \\ 1 & 0 & 0 & -1 \end{pmatrix} \begin{pmatrix} \tau_0 \\ \tau_1 \\ \tau_2 \\ \tau_3 \end{pmatrix}, \quad (24)$$

and

$$\begin{pmatrix} \rho^{\boldsymbol{\sigma}}_{+x} \\ \rho^{\boldsymbol{\sigma}}_{+y} \\ \rho^{\boldsymbol{\sigma}}_{+z} \\ \rho^{\boldsymbol{\sigma}}_{-z} \end{pmatrix} = \frac{1}{2}\begin{pmatrix} 1 & 1 & 0 & 0 \\ 1 & 0 & 1 & 0 \\ 1 & 0 & 0 & 1 \\ 1 & 0 & 0 & -1 \end{pmatrix} \begin{pmatrix} \sigma_0 \\ \sigma_1 \\ \sigma_2 \\ \sigma_3 \end{pmatrix}. \quad (25)$$

And the $B$ matrix in the main text now is the direct product of the two matrices above. And the inverse of the $B$ matrix now is given by $B^{-1} = b^{-1}\otimes b^{-1}$ where

$$b^{-1} = \begin{pmatrix} 0 & 0 & 1 & 1 \\ 2 & 0 & -1 & -1 \\ 0 & 2 & -1 & -1 \\ 0 & 0 & 1 & -1 \end{pmatrix}. \quad (26)$$

The coefficients of the WFF matrix now is given by

$$f_{ij}(\mathbf{k})A(\mathbf{k},\omega) = \sum_{\tau,\sigma}[B^{-1}]_{ij,\tau\sigma}\mathcal{I}_{\tau\sigma}, \quad (27)$$

where $\tau = LH, LV, RC, DP$ and $\sigma = +x, +y, +z, -z$.

Below we provide the detailed protocol in experiments to extract the WFF matrix elements. Suppose we firstly fix the spin to along with the $i$-direction (with algebraic sign included in $i$). It is straightforward to show that

$$\begin{aligned}
\text{Tr}[\rho_i^{\boldsymbol{\sigma}} F^{(n)}(\mathbf{k})]_{11} &= \mathcal{I}_{LH}^i(\mathbf{q}, \omega) \\
\text{Re}\left[\text{Tr}[\rho_i^{\boldsymbol{\sigma}} F^{(n)}(\mathbf{k})]_{12}\right] &= \mathcal{I}_{DP}^i(\mathbf{q}, \omega) - \mathcal{I}_{ADP}^i(\mathbf{q}, \omega) \\
\text{Im}\left[\text{Tr}[\rho_i^{\boldsymbol{\sigma}} F^{(n)}(\mathbf{k})]_{12}\right] &= \mathcal{I}_{LC}^i(\mathbf{q}, \omega) - \mathcal{I}_{RC}^i(\mathbf{q}, \omega) \\
\text{Tr}[\rho_i^{\boldsymbol{\sigma}} F^{(n)}(\mathbf{k})]_{22} &= \mathcal{I}_{LV}^i(\mathbf{q}, \omega).
\end{aligned} \tag{28}$$

Notice that it is possible to extract the polarization-relevant matrix elements by 4-independent measurements with LH, LV, DP, RC polarizations.

Then we focus on $\text{Tr}[\rho_i^{\boldsymbol{\sigma}} F^{(n)}(\mathbf{k})]_{11}$ to illustrate how to resolve the spin degree of freedom. Similar to what we've done in the light-polarization part, we have

$$\begin{aligned}
\left[F^{(n)}(\mathbf{k})\right]_{1\uparrow,1\uparrow} &= \mathcal{I}_{LH}^{+z}(\mathbf{q}, \omega) \\
\text{Re}\left[[F^{(n)}(\mathbf{k})]_{1\uparrow,1\downarrow}\right] &= \mathcal{I}_{LH}^{+x}(\mathbf{q}, \omega) - \mathcal{I}_{LH}^{-x}(\mathbf{q}, \omega); \\
\text{Im}\left[[F^{(n)}(\mathbf{k})]_{1\uparrow,1\downarrow}\right] &= \mathcal{I}_{LH}^{-y}(\mathbf{q}, \omega) - \mathcal{I}_{LH}^{+y}(\mathbf{q}, \omega); \\
\text{Im}\left[[F^{(n)}(\mathbf{k})]_{1\downarrow,1\downarrow}\right] &= \mathcal{I}_{LH}^{-z}(\mathbf{q}, \omega).
\end{aligned} \tag{29}$$

By repeating this procedure to remaining matrix elements in Eq. (28), we can obtain all the $4 \times 4$ matrix elements by at least 16 independent measurements.

Sometimes we want to measure the wavefunction information of $N_b$ different bands which may not necessarily be degenerate. The full ARPES intensity is given by

$$\mathcal{I} \propto \text{Tr}[(\rho^{\boldsymbol{\epsilon}} \otimes \rho^{\boldsymbol{\sigma}}) F^{(n)}(\mathbf{k})] \delta(\omega - \varepsilon_{n\mathbf{k}}). \tag{30}$$

Suppose we fix the scattering momentum $\mathbf{q}$, and look at the energy distribution curve (EDC). By fitting EDC and find the corresponding energies of our $N_b$ bands, it is possible to extract the projector matrix elements $[F^{(n)}(\mathbf{k})]_{ab}$ for all the $N_b$ bands by scanning the frequencies.

## II. EXTRACTION OF THE CHERN NUMBER

### 1. Non-degenerate band

In this subsection, we prove the relation

$$C = \sum_i d(\mathbf{k}_i) \tag{31}$$

for the non-degenerate electronic band. We focus on the absolute value of the inner product $|\langle u_{n\mathbf{k}} | W_{a\mathbf{q}} \rangle|^2$ with $\alpha, \sigma$ fixed. Below we set $(\alpha, \sigma) = a$ to label the probe states. Suppose on a closed manifold, which could be the Fermi surface enclosing the Weyl point, the wavefunction $|u_{n\mathbf{k}}\rangle$ has a nonzero Chern number. The proof relies on partitioning the manifold into two patches. Patch I consists of the union of $N$ small, disjoint disks, $D_i$, each centered around a node $\mathbf{k}_i$. The boundary of each disk is the loop $c_i$. Within each disk, we can adopt a smooth gauge, and let us call this gauge-$I$, and the wavefunctions in this gauge are $|u_{n\mathbf{k}}\rangle^I$. Patch II is the rest of the manifold, $S^2/(\cup_i D_i)$ and in this patch, the inner product, $\langle u_{n\mathbf{k}} | W_{a\mathbf{q}} \rangle$ is never zero. The boundary of Region II is the set of all loops $c_i$, but traversed in the opposite direction.

We observe that the gauge choice $I$ cannot be continued to Region II: because the inner product is non-zero and has non-zero winding number, it will be ill-defined at some point in this region. Thus, we need a new gauge-$II$ to untwist the phases in the wavefunction. This is achieved by defining

$$|u_{n\mathbf{k}}\rangle^{II} = \frac{{}^I\langle u_{n\mathbf{k}} | W_{a\mathbf{q}} \rangle}{|{}^I\langle u_{n\mathbf{k}} | W_{a\mathbf{q}} \rangle|} |u_{n\mathbf{k}}\rangle^I = \frac{\hat{P}_{n\mathbf{k}} |W_{a\mathbf{q}}\rangle}{\langle W_{a\mathbf{q}} | \hat{P}_{n\mathbf{k}} | W_{a\mathbf{q}} \rangle^{1/2}}. \tag{32}$$

This is possible because $|\langle u_{n\mathbf{k}}|W_{a\mathbf{q}}\rangle|$ is never zero in Region II. Since $|W_{a\mathbf{q}}\rangle$ and $\hat{P}_{n\mathbf{k}}$ are globally smooth, single-valued function, $|u_{n,\mathbf{k}}^{II}\rangle$ is also smooth and single-valued in Region II.

The Chern number is defined as the integral of the Berry curvature over the entire sphere. We can split the integral into two parts [3]:

$$\begin{aligned} C &= \frac{1}{2\pi} \int_{S^2} d^2k \mathcal{F} \\ &= \frac{1}{2\pi} (\sum_i \int_{D_i} d^2k \mathcal{F} + \int_{\text{Region II}} d^2k \mathcal{F}) \\ &= \frac{1}{2\pi} (\sum_i \oint_{c_i} d\mathbf{k} \cdot (\mathbf{A}^I - \mathbf{A}^{II})). \end{aligned} \quad (33)$$

In the last equation, we used the Stokes' theorem and used the fact that boundary of Region II is the set of all loops $c_i$, but transversed in the opposite direction. By noticing that

$$\mathbf{A}^{II} = \mathbf{A}^I - \nabla \chi(\mathbf{k}), \quad (34)$$

where $e^{i\chi(\mathbf{k})} \equiv \frac{\langle u_{n\mathbf{k}}|W_{a\mathbf{q}}\rangle}{|\langle u_{n\mathbf{k}}|W_{a\mathbf{q}}\rangle|}$. So the Chern number is given by

$$C = \frac{1}{2\pi} \sum_i \oint_{c_i} d\mathbf{k} \cdot \nabla \chi(\mathbf{k}) = \frac{1}{2\pi i} \oint_{c_i} d\log(\langle u_{n\mathbf{k}}|W_{a\mathbf{q}}\rangle) = \sum_i d_i. \quad (35)$$

### 2. Degenerate electronic bands

In this subsection, we discuss how to construct the WFF determinant from spectral functions for the doubly-degenerate electronic bands. We denote the degenerate Bloch doublet as $\{|u_{1\mathbf{k}}\rangle, |u_{2\mathbf{k}}\rangle\}$. To construct the WFF, we need a topologically trivial probe state doublet, which can be $\{|W_{(LH,\sigma)\mathbf{q}}\rangle, |W_{(LV,\sigma)\mathbf{q}}\rangle\}$ (for a fixed final state spin) or $\{|W_{(\epsilon,\uparrow)\mathbf{q}}\rangle, |W_{(\epsilon,\downarrow)\mathbf{q}}\rangle\}$ for a fixed light polarization. And the WFF determinant is given by

$$\det[F(\mathbf{k})] = \det[\langle W_{a\mathbf{q}}|\hat{P}_{\mathbf{k}}|W_{b\mathbf{q}}\rangle], \quad (36)$$

where $\hat{P}_{\mathbf{k}} = |u_{1\mathbf{k}}\rangle\langle u_{1\mathbf{k}}| + |u_{2\mathbf{k}}\rangle\langle u_{2\mathbf{k}}|$. Since we know all the matrix elements in the WFF matrix, it is straightforward to calculate $\det[F(\mathbf{k})]$.

If the probe state doublet is formed for different light polarization, the WFF determinant is given by

$$\det[F(\mathbf{k})] = \quad (37)$$
$$[\mathcal{I}_0(\mathbf{q}, \varepsilon_{n\mathbf{k}})^2 - (\mathcal{I}_{DP}(\mathbf{q}, \varepsilon_{n\mathbf{k}}) - \mathcal{I}_{ADP}(\mathbf{q}, \varepsilon_{n\mathbf{k}}))^2 - (\mathcal{I}_{RC}(\mathbf{q}, \varepsilon_{n\mathbf{k}}) - \mathcal{I}_{LC}(\mathbf{q}, \varepsilon_{n\mathbf{k}}))^2 - (\mathcal{I}_{LH}(\mathbf{q}, \varepsilon_{n\mathbf{k}}) - \mathcal{I}_{LV}(\mathbf{q}, \varepsilon_{n\mathbf{k}}))^2]/4. \quad (38)$$

where $\mathcal{I}_0(\mathbf{q}, \varepsilon_{n\mathbf{k}}) = \mathcal{I}_{LV}(\mathbf{q}, \varepsilon_{n\mathbf{k}}) + \mathcal{I}_{LH}(\mathbf{q}, \varepsilon_{n\mathbf{k}})$ and we fix the spin of the outgoing photoelectron. Also, we can also construct the probe state pair by spin-resolution and the WFF determinant is given by

$$\det[F(\mathbf{k})] = \quad (39)$$
$$[\mathcal{I}^0(\mathbf{q}, \varepsilon_{n\mathbf{k}})^2 - (\mathcal{I}^{+x}(\mathbf{q}, \varepsilon_{n\mathbf{k}}) - \mathcal{I}^{-x}(\mathbf{q}, \varepsilon_{n\mathbf{k}}))^2 - (\mathcal{I}^{+y}(\mathbf{q}, \varepsilon_{n\mathbf{k}}) - \mathcal{I}^{-y}(\mathbf{q}, \varepsilon_{n\mathbf{k}}))^2 - (\mathcal{I}^{+z}(\mathbf{q}, \varepsilon_{n\mathbf{k}}) - \mathcal{I}^{-z}(\mathbf{q}, \varepsilon_{n\mathbf{k}}))^2]/4, \quad (40)$$

where we take a fixed incident light polarization.

Like the non-degenerate band case, non-zero Chern number will lead zeros of the WFF determinant on a closed manifold. First we define $M_{na} = \langle u_{m\mathbf{k}}|W_a\rangle$ and notice that $F = M^\dagger M$. If the WFF determinant is globally non-zero on the closed manifold, we can define a new degenerate doublet

$$|\tilde{u}_{m\mathbf{k}}\rangle = \sum_n |u_{n\mathbf{k}}\rangle (MF^{-1/2})_{nm}. \quad (41)$$

The doublet is smooth, single-valued and spans 2-dimensional subspace (due the non-zero determinant), which contradicts the non-zero Chern number. So the WFF determinant must vanish at some point (spectral node) on the closed manifold.

To prove $C = \sum_i d_i$, where $d_i = \frac{1}{2\pi i} \oint_{c_i} d\mathbf{k} \det[\langle u_{m\mathbf{k}}|W_{a\mathbf{q}}\rangle]$, we again divide the closed manifold into two patches as we do in the non-degenerate band case. But the transition function between these two patches is given by

$$t(\mathbf{q}) = -\text{Tr}[U^{-1}\nabla_\mathbf{k} U] = -\nabla_\mathbf{k} \log \det(U), \tag{42}$$

where $U = MF^{-1/2}$ is a unitary matrix. By applying the Stokes' theorem, we have

$$C = \frac{1}{2\pi i} \sum_i \oint_{c_i} d\mathbf{k} \cdot \nabla_\mathbf{k} \log \det(U) = \sum_i d_i, \tag{43}$$

and here we use the fact that $F^{-1/2}$ is a Hermitian matrix and will not contribute to the phase winding.

## III. EXTRACTION OF THE BERRY PHASE AND $Z_2$ MONOPOLE CHARGE

In this section we consider how to extract the topological invariants in a system with composite inversion $P$ and time-reversal $T$ symmetry, with $(PT)^2 = 1$. The generic band crossing is nodal line, characterized by two $Z_2$ indices: Berry phase (single band) and $Z_2$ monopole charge (multi-band). The discussion about the Berry phase and $Z_2$ monopole charge is general and can be directly applied to phonon and magnon bands.

*Berry phase*—Since $(PT)^2 = 1$, one can find a basis such that the wavefunctions $|u_{n\mathbf{k}}\rangle$ are locally real. The Berry phase of the wavefunction is defined as the loop integral of the Berry connection on the loop linking the nodal line

$$\gamma_n = \oint_c d\mathbf{k} \cdot \mathbf{A}_n(\mathbf{k}) = -i \oint d\mathbf{k} \cdot \langle u_{n\mathbf{k}}|\nabla_\mathbf{k}|u_{n\mathbf{k}}\rangle. \tag{44}$$

The $\pi$-Berry phase is an obstruction to finding a globally smooth, real gauge for the wavefunction $|u_{n\mathbf{k}}\rangle$. So the overlap between $|u_{n\mathbf{k}}\rangle$ and a $PT$-symmetry probe state $|W\rangle$ must have a node on the loop. If not, we can do a gauge transformation to the wavefunction

$$|u_{n\mathbf{k}}\rangle' = \hat{P}_{n\mathbf{k}}|W\rangle/(\langle W|\hat{P}_\mathbf{k}|W\rangle)^{1/2}, \tag{45}$$

which is real and globally smooth on the linking loop. Notice that the number of nodes on the loop must be odd, because the zeros can merge in pairs under smooth deformation. Under smooth deformation of the linking loop, the zeros will form a membrane bounded by the nodal line.

In ARPES, the probe states are indeed $PT$-symmetric. Suppose we work at the length gauge, where the probe states are $|W_{\alpha\mathbf{q}}\rangle = e^{-i\mathbf{k}\cdot\hat{\mathbf{r}}}\hat{r}_\alpha|\chi_\mathbf{q}\rangle$, and we ignore the spin indices due to $(PT)^2 = 1$. It is a good approximation to take $|\chi_\mathbf{q}\rangle$ as a plane wave state, and we have $PT|W_{\alpha\mathbf{q}}\rangle = |W_{\alpha\mathbf{q}}\rangle$.

$Z_2$ *monopole charge*— The $Z_2$ monopole charge is an obstruction to defining a globally smooth real gauge on the sphere enclosing the nodal line. This monopole charge can only be nontrivial for multi-band systems. Like the Chern number of the degenerate band case, the nontrivial $Z_2$ monopole charge will lead to zeros of $\det F$ on the sphere, or one can use the probe states to define a globally smooth real gauge on the sphere.

But different than the Chern number case, due to $(PT)^2 = 1$, the WFF matrix $F_{ab}(\mathbf{k})$ is locally real since $\{|u_{n\mathbf{k}}\rangle\}$ are locally real and $|W_{a\mathbf{q}}\rangle$ are $PT$-symmetric. Suppose we use $\theta, \phi$ to parametrize the sphere, and the zero condition is

$$\det F(\theta, \phi) = 0. \tag{46}$$

There is one constraint but two variables, so the allowed solutions will form a 1D loop on the sphere. One can think the loop is the domain wall that separates the regions where $\det F$ is negative or positive. The loop will trace out a nodal tube connecting nodal lines with opposite $Z_2$ monopole charges. The nodal loop can shrink to a nodal arc under certain symmetry constraint.

## IV. TOPOLOGICAL TEXTURE

Within the "Topological selection rule" framework, not only can we construct a WFF determinant that manifest zeros protected by the topological charge of the band crossing, but also directly visualize the wavefunction textures.

To illustrate, let us first consider a non-degenerate electronic band. We define the two-component spinor $s_{\mathbf{k}}$:

$$s_{\mathbf{k}} = \frac{1}{(|a_{\mathbf{k}}|^2 + |b_{\mathbf{k}}|^2)^{1/2}} \begin{pmatrix} a_{\mathbf{k}} \\ b_{\mathbf{k}} \end{pmatrix}, \tag{47}$$

where $a_{\mathbf{k}} = \langle W_1 | u_{\mathbf{k}} \rangle$, $b_{\mathbf{k}} = \langle W_2 | u_{\mathbf{k}} \rangle$. We suppose this spinor is never a null vector on the sphere, which is guaranteed by the linear-independence of the probes. Then we show this newly defined spinor has the same topology of the original Bloch wave function $|\psi_{n\mathbf{k}}\rangle$.

Consider partitioning the sphere into two patches. On patch I, $|a_{\mathbf{k}}| \neq 0$ and on patch II, $|b_{\mathbf{k}}| \neq 0$. On patch I, we define the gauge

$$s_{\mathbf{k}}{}^{I} = \frac{1}{(1+|\zeta^{-1}|^2)} \begin{pmatrix} 1 \\ \zeta^{-1} \end{pmatrix}, \tag{48}$$

and on patch II, we have

$$s_{\mathbf{k}}{}^{II} = \frac{1}{(1+|\zeta|^2)} \begin{pmatrix} \zeta \\ 1 \end{pmatrix}, \tag{49}$$

where $\zeta = a_{\mathbf{k}}/b_{\mathbf{k}}$.

On the overlap of patch I and patch II, which we dub the "equator", the transition function is given by

$$s_{\mathbf{k}}{}^{II} = e^{i \arg \zeta} s_{\mathbf{k}}{}^{I}. \tag{50}$$

From Stokes' theorem, we know the Chern number of this spinor is given by

$$C = \oint_{\partial \Sigma_{II}} d \arg \zeta. \tag{51}$$

Note that $\partial \Sigma_{II}$ is exactly the equator but with orientation defined by the patch II.

We now use this gauge fixing method to calculate the Chern number of the original wavefunction. On patch I, since $a_{\mathbf{k}} = \langle W_1 | \psi_{\mathbf{k}} \rangle$ is never zero we can fix the phase of $|u_{\mathbf{k}}\rangle$ such that

$$|u_{\mathbf{k}}\rangle^{I} = \frac{\langle u_{\mathbf{k}} | W_1 \rangle}{|\langle u_{\mathbf{k}} | W_1 \rangle|} |u_{\mathbf{k}}\rangle = e^{-i \arg a_{\mathbf{k}}} |u_{\mathbf{k}}\rangle, \tag{52}$$

such that under this gauge choice $\langle W_1 | u_{\mathbf{k}} \rangle^{I}$ is always real and positive. Similarly on the patch II, we choose the gauge

$$|u_{\mathbf{k}}\rangle^{II} = \frac{\langle u_{\mathbf{k}} | W_2 \rangle}{|\langle u_{\mathbf{k}} | W_2 \rangle|} |u_{\mathbf{k}}\rangle = e^{-i \arg b_{\mathbf{k}}} |\psi_{\mathbf{k}}\rangle. \tag{53}$$

On the equator, the transition function is given by

$$|\psi_{\mathbf{k}}\rangle^{II} = |\psi_{\mathbf{k}}\rangle^{I} e^{-i(\arg b_{\mathbf{k}} - \arg a_{\mathbf{k}})} = |\psi_{\mathbf{k}}\rangle^{I} e^{i \arg \zeta}. \tag{54}$$

It has the same transition function as the two-component spinor defined above, which means the Chern number of these complex vectors must be the same.

We know for a 2-component spinor, the Chern number is equal to the wrapping number of the pseudo-spin $\mathbf{S}(\mathbf{k})$:

$$\mathbf{S}(\mathbf{k}) = \sum_{ij} s_i^*(\mathbf{k}) \boldsymbol{\sigma}_{ij} s_j(\mathbf{k}). \tag{55}$$

More importantly, the pseudo-spin can be constructed from the ARPES intensities

$$\begin{aligned} S_x(\mathbf{k}) &= \frac{\mathcal{I}_{DP} - \mathcal{I}_{ADP}}{\mathcal{I}_{LH} + \mathcal{I}_{LV}}; \\ S_y(\mathbf{k}) &= \frac{\mathcal{I}_{RC} - \mathcal{I}_{LC}}{\mathcal{I}_{LH} + \mathcal{I}_{LV}}; \\ S_z(\mathbf{k}) &= \frac{\mathcal{I}_{LH} - \mathcal{I}_{LV}}{\mathcal{I}_{LH} + \mathcal{I}_{LV}}. \end{aligned} \tag{56}$$

Above, we construct the pseudo-spin from light dichroism. Also, one can construct $\mathbf{S}(\mathbf{k})$ from "spin dichroism". With a fixed incident light polarization, we have

$$
\begin{aligned}
S_x(\mathbf{k}) &= \frac{\mathcal{I}^{+x} - \mathcal{I}^{-x}}{\mathcal{I}^{+z} + \mathcal{I}^{-z}}; \\
S_y(\mathbf{k}) &= \frac{\mathcal{I}^{+y} - \mathcal{I}^{-y}}{\mathcal{I}^{+z} + \mathcal{I}^{-z}}; \\
S_z(\mathbf{k}) &= \frac{\mathcal{I}^{+z} - \mathcal{I}^{-z}}{\mathcal{I}^{+z} + \mathcal{I}^{-z}}.
\end{aligned}
\tag{57}
$$

It is straightforward to generalize it to a doubly-degenerate band by choosing two sets of linearly independent probe state pairs $|W_{1/2}\rangle, |\tilde{W}_{1/2}\rangle$ Now the component of the spinor is defined as

$$
\begin{aligned}
a_{\mathbf{k}} &= \det \langle W_i | u_j \rangle; \\
b_{\mathbf{k}} &= \det \langle \tilde{W}_i | u_j \rangle.
\end{aligned}
\tag{58}
$$

In the ARPES's context, $|W\rangle$ can be photoelectrons with spin polarized along $+z$ with different incident light polarization and $|\tilde{W}\rangle$ can be photoelectrons with spin polarized along $-z$.

Like we have done in the non-degenerate case, the Chern number of the two-component spinor $s_{\mathbf{k}}$ is given by

$$
C = \oint_{\Sigma_{II}} d\arg \zeta. \tag{59}
$$

It is straightforward to see this Chern number equals the Chern number of the Bloch wave function doublet. On patch I, we choose the determinant $\det[\langle W_i | u_j \rangle^I]$ to be real and positive and we have

$$
|u_i\rangle^I = \sum_j |u_j\rangle \langle u_j | W_i\rangle. \tag{60}
$$

The orthonormalized wavefunctions are given by $|u_i\rangle^I = \sum_j |u_j\rangle U_{ji}$, where $U = MF^{-1/2}$, where $M_{ij} = \langle u_i | W_j \rangle$ and $F = M^\dagger M$. And on patch II we have

$$
|u_i\rangle^{II} = \sum_j |u_j\rangle \langle u_j | \tilde{W}_i \rangle, \tag{61}
$$

or $|u_i\rangle^{II} = \sum_j V_{ij} |u_i\rangle$ where $V = M'(M'^\dagger M')^{-1/2}$ and $M' = \langle u_i | \tilde{W}_j \rangle$. By applying the Stokes' theorem to the non-Abelian version of the Chern number formula, we have

$$
C = \frac{i}{2\pi} \oint_{\Sigma_{II}} d\log(\det(VU^{-1})) = \oint_{\Sigma_{II}} \arg \zeta. \tag{62}
$$

This proves the Bloch wave function and the two component spinor has the same Chern number.

Now we explicitly write out the pseudo-spin quantity in this case.

$$
\begin{aligned}
S_x(\mathbf{k}) &= \frac{1}{|a|^2+|b|^2}(a^*b + ab^*) = \frac{1}{|\det \langle W_i|u_j\rangle|^2 + |\det \langle \tilde{W}_i|u_j\rangle|^2}(\det \langle u_i|W_j\rangle \det \langle \tilde{W}_i|u_j\rangle + c.c.) \\
S_y(\mathbf{k}) &= \frac{i}{|a|^2+|b|^2}(a^*b - ab^*) = \frac{i}{|\det \langle W_i|u_j\rangle|^2 + |\det \langle \tilde{W}_i|u_j\rangle|^2}(\det \langle u_i|W_j\rangle \det \langle \tilde{W}_i|u_j\rangle - c.c.) \\
S_z(\mathbf{k}) &= \frac{1}{|a|^2+|b|^2}(|a|^2 - |b|^2) = \frac{1}{|\det \langle W_i|u_j\rangle|^2 + |\det \langle \tilde{W}_i|u_j\rangle|^2}(|\det \langle W_i|u_j\rangle|^2 - |\det \langle \tilde{W}_i|u_j\rangle|^2).
\end{aligned}
\tag{63}
$$

In realistic experiments, we can choose such reference states

$$
\begin{aligned}
|W_{1/2}\rangle &= e^{-i\mathbf{k}\cdot\hat{\mathbf{r}}} \hat{v}_{x/y} |\chi_{\mathbf{q}\uparrow}\rangle \\
|\tilde{W}_{1/2}\rangle &= e^{-i\mathbf{k}\cdot\hat{\mathbf{r}}} \hat{v}_{x/y} |\chi_{\mathbf{q}\downarrow}\rangle.
\end{aligned}
\tag{64}
$$

And we define an intensity triplet:

$$\begin{aligned}\mathcal{I}^{\uparrow\downarrow}(\epsilon) &= \frac{1}{2}\left[(\mathcal{I}^{+x}(\epsilon) - i\mathcal{I}^{+y}(\epsilon)) - (\mathcal{I}^{-x}(\epsilon) - i\mathcal{I}^{-y}(\epsilon))\right] = \langle\uparrow,\epsilon|\hat{P}|\downarrow,\epsilon\rangle \\ \mathcal{I}^{+z}(\epsilon) &= \langle\uparrow,\epsilon|\hat{P}|\uparrow,\epsilon\rangle \\ \mathcal{I}^{-z}(\epsilon) &= \langle\downarrow,\epsilon|\hat{P}|\downarrow,\epsilon\rangle.\end{aligned} \tag{65}$$

A few things to notice. First, $|\uparrow,\epsilon\rangle = \epsilon \cdot \hat{\mathbf{v}}|\chi_{\mathbf{q}\uparrow}\rangle$. Second, $I^{\uparrow\downarrow}(\epsilon)$ is a complex number.

Then we define a function to extract the squared determinant information:

$$g(\mathcal{I}^s) = \frac{1}{4}\left[(\mathcal{I}^s_{LH} + \mathcal{I}^s_{LV})^2 - ((\mathcal{I}^s_{DP} - \mathcal{I}^s_{ADP})^2 + (\mathcal{I}^s_{RC} - \mathcal{I}^s_{LC})^2 + (\mathcal{I}^s_{LH} - \mathcal{I}^s_{LV})^2)\right], \tag{66}$$

with $s$ being $\uparrow\downarrow, \pm z$. So the pseudo-spin can be expressed as

$$\begin{aligned}S_x(\mathbf{k}) &= \frac{g(\mathcal{I}^{\uparrow\downarrow}) + g^*(\mathcal{I}^{\uparrow\downarrow})}{g(\mathcal{I}^{+z}) + g(\mathcal{I}^{-z})} \\ S_y(\mathbf{k}) &= i\frac{g(\mathcal{I}^{\uparrow\downarrow}) - g^*(\mathcal{I}^{\uparrow\downarrow})}{g(\mathcal{I}^{+z}) + g(\mathcal{I}^{-z})} \\ S_z(\mathbf{k}) &= \frac{g(\mathcal{I}^{+z}) - g(\mathcal{I}^{-z})}{g(\mathcal{I}^{+z}) + g(\mathcal{I}^{-z})}\end{aligned} \tag{67}$$

## V. THE WILSON LOOP AND NON-ABELIAN BERRY CURVATURE

In this section, we show how to extract the spectrum of the Wilson loop operator and the non-Abelian Berry curvature by ARPES measurements. With the four independent probe states $|W_{(\alpha,\sigma)}\rangle$, we can define a projected Bloch doublet

$$\psi_{1/2} = \begin{pmatrix} \langle W_{(x,\uparrow)}|u_{1/2\mathbf{k}}\rangle \\ \langle W_{(y,\uparrow)}|u_{1/2\mathbf{k}}\rangle \\ \langle W_{(x,\downarrow)}|u_{1/2\mathbf{k}}\rangle \\ \langle W_{(y,\downarrow)}|u_{1/2\mathbf{k}}\rangle \end{pmatrix}. \tag{68}$$

And we define a $4 \times 2$ complex matrix $\Psi = (\psi_1, \psi_2)$. Below we denote

$$[F_{n\mathbf{k}}]_{ab} = \langle W_a|\hat{P}_{n\mathbf{k}}|W_b\rangle, \tag{69}$$

where $a, b = (\alpha, \sigma) = 1, 2, 3, 4$ and we have $F_{n\mathbf{k}} = \Psi\Psi^\dagger$. But the matrix $F_{n\mathbf{k}}$ is not a projector: since the projected states $\psi_{1/2}$ are not orthonormalized ($\Psi^\dagger\Psi \neq I$), $F_{n\mathbf{k}}^2 \neq F_{n\mathbf{k}}$. The orthonormalized projected states are given by

$$\tilde{\Psi} = \Psi(\Psi^\dagger\Psi)^{-1/2}, \tag{70}$$

and we have $(\tilde{\Psi}\tilde{\Psi}^\dagger)^2 = \tilde{\Psi}\tilde{\Psi}^\dagger$. One problem is that we have no access to the $2 \times 2$ matrix $\Psi^\dagger\Psi$ in ARPES experiments, so we need an alternative route to construct the projectors. Since we know each matrix element in $F_{n\mathbf{k}} = \Psi\Psi^\dagger$, we can perform the spectral decomposition:

$$[\hat{P}_{n\mathbf{k}}] = U\begin{pmatrix} \sigma_1 & & & \\ & \sigma_2 & & \\ & & 0 & \\ & & & 0 \end{pmatrix}U^\dagger, \tag{71}$$

where $\sigma_{1/2}$ are non-negative real numbers. The projector can be obtained by "flattening" the $G$ matrix

$$\Pi = U\begin{pmatrix} 1 & & & \\ & 1 & & \\ & & 0 & \\ & & & 0 \end{pmatrix}U^\dagger, \tag{72}$$

and it is straightforward to check $\Pi^2 = \Pi$. Also we have $\tilde{\Psi}\tilde{\Psi}^\dagger = \Pi$.

The Wilson loop operator, by definition, is given by the ordered product

$$\mathcal{W}_\mathcal{C} = \prod_{\mathbf{k_i}}^{\mathcal{C}} \Pi(\mathbf{k}_i), \tag{73}$$

and the matrix elements are given by

$$\langle \tilde{u}_i | \mathcal{W}_\mathcal{C} | \tilde{u}_j \rangle = \tilde{u}_i^\dagger(\mathbf{k}_0)(\prod_{\mathbf{k}_i}^{\mathcal{C}} \Pi(\mathbf{k}_i))\tilde{u}_j(\mathbf{k}_0) \\ = \tilde{u}_i^\dagger(\mathbf{k}_0)(\Pi(\mathbf{k}_0)\Pi(\mathbf{k}_N)\Pi(\mathbf{k}_{N-1})\ldots\Pi(\mathbf{k}_1)\Pi(\mathbf{k}_0))\tilde{u}_j(\mathbf{k}_0). \tag{74}$$

We know the spectrum of the Wilson loop operator is a gauge invariant quantity. Notice that

$$\langle \tilde{u}_i | \mathcal{W}_\mathcal{C} | \tilde{u}_j \rangle = [\Psi^\dagger(\mathbf{k}_0)(\Pi(\mathbf{k}_0)\Pi(\mathbf{k}_N)\Pi(\mathbf{k}_{N-1})\ldots\Pi(\mathbf{k}_1)\Pi(\mathbf{k}_0))\Psi(\mathbf{k}_0)]_{ij}. \tag{75}$$

We have the lemma: the matrix $AB$ and $BA$ have the same non-zero spectrum. Then the non-zero spectrum of the Wilson loop operator is identical to the matrix

$$\Pi(\mathbf{k}_0)\Pi(\mathbf{k}_N)\Pi(\mathbf{k}_{N-1})\ldots\Pi(\mathbf{k}_1)\Pi(\mathbf{k}_0), \tag{76}$$

which can be obtained from ARPES measurements.

If we consider the loop traversing the whole Brillouin zone along a primitive reciprocal lattice vector, the Wilson loop operator is given by

$$\mathcal{W}_\mathbf{g}(\mathbf{k}_\perp) = \lim_{N\to\infty} \Pi(\mathbf{k}_\perp + \mathbf{g})\Pi(\mathbf{k}_\perp + \frac{N-1}{N}\mathbf{g})\Pi(\mathbf{k}_\perp + \frac{N-2}{N}\mathbf{g})\ldots\Pi(\mathbf{k}_\perp + \frac{1}{N}\mathbf{g})\Pi(\mathbf{k}_\perp). \tag{77}$$

The matrix elements are given by $\langle \tilde{u}_{i\mathbf{k}_\perp+\mathbf{g}}|\mathcal{W}_\mathbf{g}(\mathbf{k}_\perp)|\tilde{u}_{j\mathbf{k}_\perp}\rangle$, and it has the same spectrum as

$$V(\mathbf{k}_\perp)\lim_{N\to\infty} \Pi(\mathbf{k}_\perp + \mathbf{g})\Pi(\mathbf{k}_\perp + \frac{N-1}{N}\mathbf{g})\Pi(\mathbf{k}_\perp + \frac{N-2}{N}\mathbf{g})\ldots\Pi(\mathbf{k}_\perp + \frac{1}{N}\mathbf{g})\Pi(\mathbf{k}_\perp), \tag{78}$$

where $V(\mathbf{k}_\perp)$ is the possible sewing matrix (depending on the gauge choice).

We know the Berry curvature is the infinitesimal generator of the holonomy (Wilson loop). By the Ambrose-Singer Theorem, we have

$$i\Omega_{nm}^{12}\delta k_1 \delta k_2 = \log W_4 W_3 W_2 W_1, \tag{79}$$

the Berry curvature times the area element equals the log of the Wilson loop on the rectangle. So one can solve the spectrum of the non-Abelian Berry curvature by constructing the Wilson loop.

We then calculate the Wilson loop spectrum in the Bernevig-Hughes-Zhang model [3], which is a model describing the $Z_2$ topological insulator. The $6\times 6$ Hamiltonian is given by

$$\begin{pmatrix} M(\mathbf{k}) & d_1(\mathbf{k}) + id_2(\mathbf{k}) & & & & \\ d_1(\mathbf{k}) - id_2(\mathbf{k}) & -M(\mathbf{k}) & & & & \\ & & \varepsilon_t(\mathbf{k}) & & & \\ & & & M(\mathbf{k}) & -d_1(\mathbf{k}) + id_2(\mathbf{k}) & \\ & & & -d_1(\mathbf{k}) + id_2(\mathbf{k}) & -M(\mathbf{k}) & \\ & & & & & \varepsilon_t(\mathbf{k}) \end{pmatrix}, \tag{80}$$

where $d_1(\mathbf{k}) = A\sin k_x$, $d_2(\mathbf{k}) = -A\sin k_y$, $M(\mathbf{k}) = M - 2B(2 - \cos k_x - \cos k_y)$. The basis are arranged as $|s,\uparrow\rangle, |p,\uparrow\rangle, |t,\uparrow\rangle, |s,\downarrow\rangle, |p,\downarrow\rangle, |t,\downarrow\rangle$, where $|t,\uparrow/\downarrow\rangle$ stands for the topologically trivial states. We choose the probe states to be

$$|W_{(\alpha,\sigma),\mathbf{k}}\rangle = \partial_\alpha H(\mathbf{k})|\tau_\sigma\rangle, \tag{81}$$

where $|\tau_\uparrow\rangle = (1,1,1,0,0,0)^T$ (similar for $|\tau_\downarrow\rangle$). Then we need to check if the effective projector $\Pi_\mathbf{k}$ has the time-reversal symmetry. For the current basis choice, the time-reversal acts as

$$T = (i\sigma_y \otimes I_{3\times 3})K \equiv U_T K, \tag{82}$$

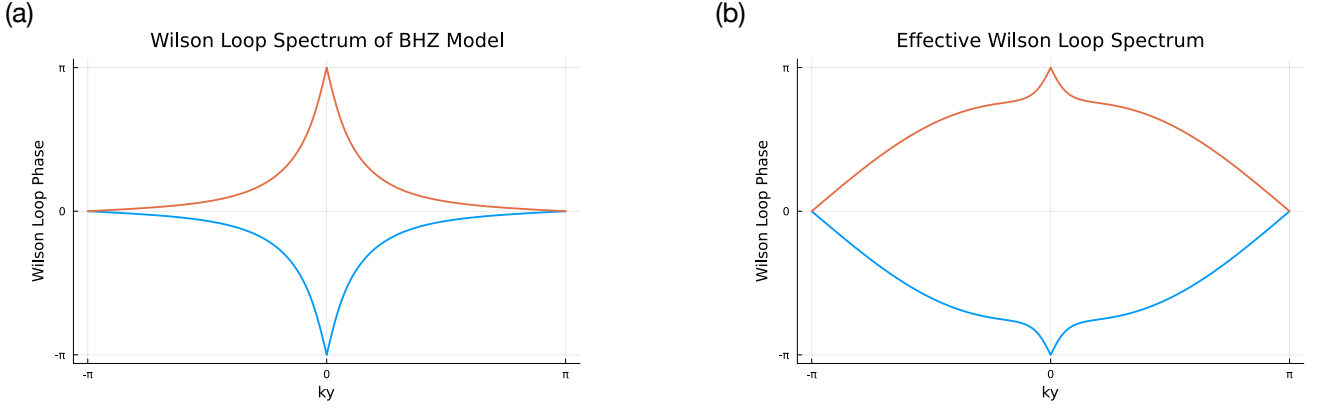

FIG. 1. We plot the Wilson loop spectra for both the band projector and the effective projector. The parameter in the Hamiltonian is chosen to be $A = -3.42, B = -1.69, M = -0.689$, and the corresponding band projector contains the Bloch states of the third and fourth band.

where $K$ is the complex conjugate, and $U_T$ is a unitary matrix, and $U_T^* U_T = -1$. Under the time-reversal symmetry, the projector matrix will transform as

$$P_{\mathbf{k}} = U_T P^*_{-\mathbf{k}} U_T^\dagger. \tag{83}$$

And for the probe states we have

$$T \partial_\alpha H(\mathbf{k}) |\tau_\sigma\rangle = -\partial_\alpha H(-\mathbf{k}) U_T K |\tau_\sigma\rangle = -\partial_\alpha H(-\mathbf{k}) |\tau_{\sigma'}\rangle (i\sigma_y)_{\sigma'\sigma}, \tag{84}$$

which is equivalent to $T |W_{(\alpha,\sigma)\mathbf{k}}\rangle = - |W_{(\alpha,\sigma')}\rangle (i\sigma_y)_{\sigma'\sigma}$. The WFF matrix is

$$F_{ab}(\mathbf{k}) = \langle W_{(\alpha,\sigma)\mathbf{k}} | P_{\mathbf{k}} | W_{(\beta,\sigma')\mathbf{k}} \rangle, \tag{85}$$

and due to the TR we have $(i\sigma_y)^\dagger F(-\mathbf{k})(i\sigma_y) = F^*(\mathbf{k})$, and if we identify $i\sigma_y$ as the time-reversal operation in the probe subspace, we reach the conclusion that $\Pi_{\mathbf{k}}$ (flattened WFF) is in the same symmetry class of $P_{\mathbf{k}}$, and $\Pi_{\mathbf{k}}$ should also have the nontrivial $Z_2$ invariant. We validate this by calculating the Wilson loop spectra, see Fig. 1.

## VI. BUNDLE ISOMORPHISM

In this section we show the mapping from the original Bloch states to the projected Bloch states is actually a "bundle isomorphism". Suppose the original Bloch states live in the Hilbert space $\mathcal{H} \simeq \mathbb{C}^N$. And the Bloch bundle $E_{\text{Bloch}} \to \mathcal{M}$ with each fiber being $E_{\mathbf{k}} = \text{span}\,\hat{P}_{\mathbf{k}}$ where $\hat{P}_{\mathbf{k}} = \sum_{n=1}^{N_b} |u_{n\mathbf{k}}\rangle \langle u_{n\mathbf{k}}|$ is the rank-$N_b$ band projector, and $\mathcal{M}$ is the base manifold which we choose as a closed one. And we have $N_p > N_b$ probe states $\{|W_{a\mathbf{q}}\rangle\}$, which is topologically trivial. And we define

$$W = (|W_{1\mathbf{q}}\rangle, \ldots, |W_{N_p\mathbf{q}}\rangle) \in \mathbb{C}^{N \times N_p}; \quad \Psi = (|u_{1\mathbf{k}}\rangle, \ldots, |u_{N_p\mathbf{k}}\rangle) \in \mathbb{C}^{N \times N_b}, \tag{86}$$

and

$$M_{an}(\mathbf{k}) = W^\dagger \Psi = \langle W_{a\mathbf{q}} | u_{n\mathbf{k}} \rangle. \tag{87}$$

We see each column of $M$ is the 4-dimensional projected Bloch wavefunction. But $MM^\dagger$ does not give the corresponding projector because the columns are not orthonormal to each other. The correct projector is given by

$$\Pi_{\mathbf{k}} = M S^{-1} M^\dagger, \quad S = M^\dagger M. \tag{88}$$

Now we define the target bundle $E_{\text{tar}} \to \mathcal{M}$, with each fiber $E_{\text{tar},\mathbf{k}}$ being span $\Pi_{\mathbf{k}}$, and the Hilbert space on each fiber $\mathcal{H}_{\text{tar}} = \mathbb{C}^{N_p}$.

First, we establish that $V(\mathbf{k})$ is an isometry on the fibers. For any two sections $|\phi_1\rangle, |\phi_2\rangle \in E_{\mathbf{k}}$, the inner product in the target bundle is given by:

$$(V(\mathbf{k})|\phi_1\rangle, V(\mathbf{k})|\phi_2\rangle) = \langle\phi_1|V^\dagger V|\phi_2\rangle \tag{89}$$

$$= \langle\phi_1|\Psi S^{-1/2} M^\dagger M S^{-1/2} \Psi^\dagger|\phi_2\rangle \tag{90}$$

$$= \langle\phi_1|\Psi S^{-1/2} S S^{-1/2} \Psi^\dagger|\phi_2\rangle \tag{91}$$

$$= \langle\phi_1|\Psi\Psi^\dagger|\phi_2\rangle. \tag{92}$$

Recalling that $\Psi\Psi^\dagger = \hat{P}_{\mathbf{k}}$ is the projector onto $E_{\mathbf{k}}$, and since $|\phi_{1,2}\rangle \in E_{\mathbf{k}}$ (implying $\hat{P}_{\mathbf{k}}|\phi\rangle = |\phi\rangle$), we obtain:

$$(V(\mathbf{k})|\phi_1\rangle, V(\mathbf{k})|\phi_2\rangle) = \langle\phi_1|\phi_2\rangle. \tag{93}$$

Thus, $V(\mathbf{k})$ preserves the inner product.

This isometry property immediately implies injectivity: if $V(\mathbf{k})|\phi\rangle = 0$, then $\langle\phi|\phi\rangle = |V(\mathbf{k})|\phi\rangle|^2 = 0$, implying $|\phi\rangle = 0$.

To prove surjectivity onto the target fiber $E_{\text{tar},\mathbf{k}}$, we compute the projector associated with the range of $V$. Using the orthonormality of the Bloch basis ($\Psi^\dagger\Psi = I_{N_b}$), we have:

$$V(\mathbf{k})V^\dagger(\mathbf{k}) = MS^{-1/2}\Psi^\dagger\Psi S^{-1/2}M^\dagger = MS^{-1}M^\dagger. \tag{94}$$

This is precisely the definition of the target effective projector $\Pi_{\mathbf{k}}$ derived in the previous section. Since $VV^\dagger = \Pi_{\mathbf{k}}$ acts as the identity on the target fiber $E_{\text{tar},\mathbf{k}}$, for any vector $w \in E_{\text{tar},\mathbf{k}}$, we can write $w = \Pi_{\mathbf{k}}w = V(V^\dagger w)$. Since $V^\dagger w \in E_{\mathbf{k}}$, every element in the target fiber is in the image of $V$, proving surjectivity.

Finally, $V(\mathbf{k})$ is smooth with respect to $\mathbf{k}$ because the constituents $\Psi(\mathbf{k})$ and $M(\mathbf{k})$ are smooth, and $S(\mathbf{k})$ is invertible (guaranteed by the no-rank-drop condition). Since $V(\mathbf{k})$ is a smooth, bijective linear map between fibers that preserves the metric, it constitutes a vector bundle isomorphism between $E_{\text{Bloch}}$ and $E_{\text{tar}}$.

## VII. QUANTUM GEOMETRY TENSOR

The quantum geometry tensor (QGT) $Q_{ij}$ is defined as

$$Q_{ij}^{nn'}(\mathbf{k}) = \langle u_{n\mathbf{k}}|\partial_i \hat{P}_{\mathbf{k}} \partial_j \hat{P}_{\mathbf{k}}|u_{n'\mathbf{k}}\rangle, \tag{95}$$

which can be decomposed as $Q_{ij} = g_{ij} - \frac{i}{2}\Omega_{ij}$, where $g_{ij} = (Q_{ij} + Q_{ij}^\dagger)/2$ is the quantum metric tensor and $\Omega_{ij} = i(Q_{ij} - Q_{ij}^\dagger)$ is the non-Abelian Berry curvature. The projector of the projected Bloch state is given by $\Pi_{\mathbf{k}}$ which is obtained by flattening the matrix $\langle W_a|\hat{P}_{\mathbf{k}}|W_b\rangle$, and one can calculate the QGT corresponding to $\Pi_{\mathbf{k}}$. But generally, it is not identical to the QGT of the original Bloch states $\hat{P}_{\mathbf{k}}$ because some geometric information gets lost in the projection procedure.

Actually, one can construct a model, where the Berry curvature of $\hat{P}_{\mathbf{k}}$ and $\Pi_{\mathbf{k}}$ are locally different, but their integrals over the Brillouin zone (the Chern number) are the same We construct a four band Weyl semimetal Hamiltonian, which reads

$$H(\mathbf{k}) = t\sum_{i=x,y,z}\sin k_i \Gamma_i + (m - \sum_{i=x,y,z}\cos k_i)\Gamma_4 + b\mathbb{1}\otimes\sigma_z, \tag{96}$$

where $\Gamma_1 = \tau_x\sigma_x, \Gamma_2 = \tau_x\sigma_y, \Gamma_3 = \tau_x\sigma_z, \Gamma_4 = \mathbb{1}\otimes\sigma_z$. We choose $t = 1, m = 2.5, b = 1.0$, such that the Weyl points are only on the $k_z$ axis. One can check the Weyl point now are given by $\mathbf{k}_W = (0, 0, \arccos 0.25)$. One the planes that lie between the Weyl points, the wavefunction will have nonzero Chern number. In Fig. 2 (a1/b1), we plot the Berry curvature and clearly they are locally different, but based on the Wilson loop spectra Fig. 2 (a2/b2), these two projectors have the same Chern number.

But there is a physical scenario where the QGT of $\Pi_{\mathbf{k}}$ can be used to approximate the one of $\hat{P}_{\mathbf{k}}$. Suppose there are four bands energetically separated from the rest of the bands and these four bands are topologically trivial. This enables us to find a tight-binding model to reproduce these four bands

$$|u_{n\mathbf{k}}\rangle = \sum_\alpha u_{n\mathbf{k}}^\alpha |\alpha\mathbf{k}\rangle, \tag{97}$$

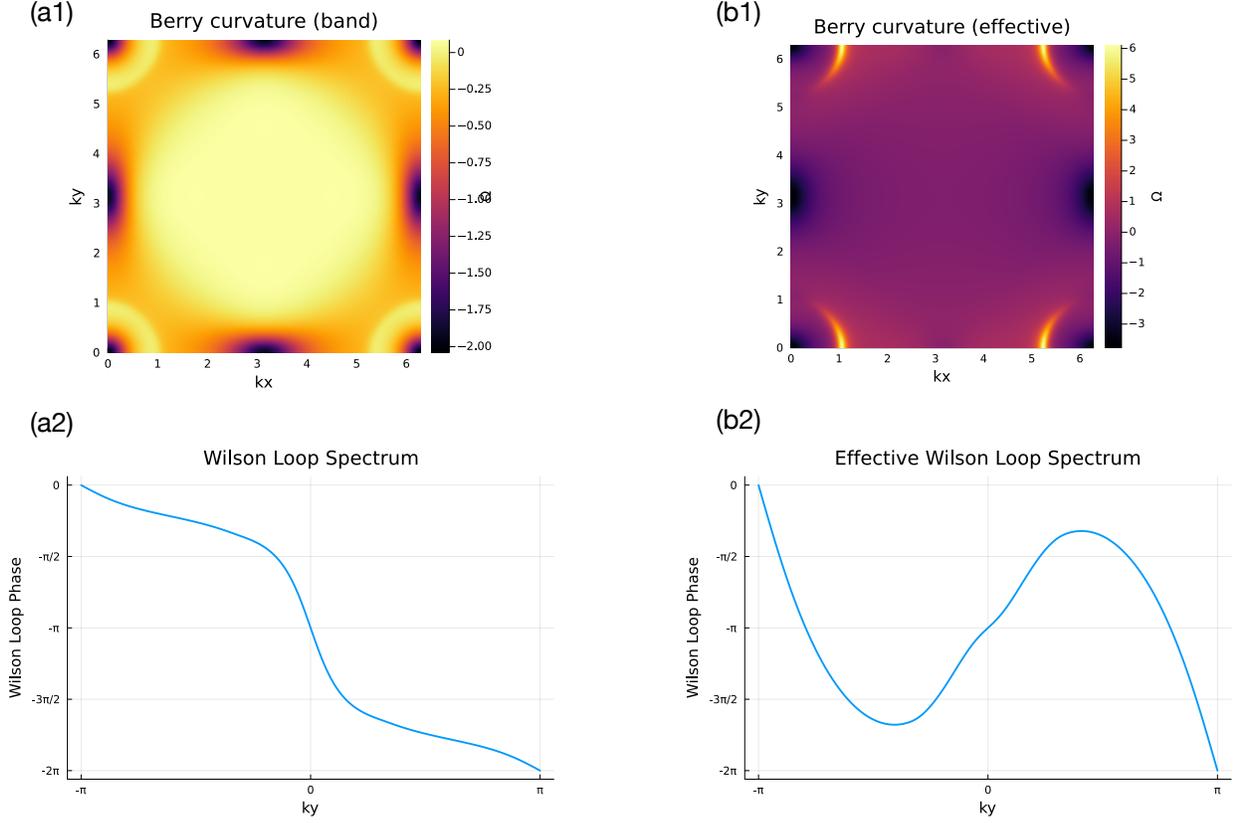

FIG. 2. We plot the Berry curvature of the band projector and the effective projector, and show despite the Berry curvatures being locally different, their Wilson loop spectra give the same charge pumping, and their Chern numbers are the same.

where $|\alpha \mathbf{k}\rangle$ is the four-dimensional tight-binding basis in the momentum-space. Suppose we have four independent probes $|W_{a\mathbf{q}}\rangle$, and we can project these probes onto the tight-binding basis

$$|\phi_{a\mathbf{k}}\rangle = \sum_\alpha |\alpha \mathbf{k}\rangle \langle \alpha \mathbf{k}|W_{a\mathbf{q}}\rangle, \tag{98}$$

so the projected probes can be represented by four independent 4D vectors with the coefficients given by $\phi_{a\mathbf{k}}^\alpha = \langle \alpha \mathbf{k}|W_{a\mathbf{q}}\rangle$, which also span the whole 4D space. We below denote the projector of the projected probe states as

$$P_{\text{probe}} = \sum_a \phi_{a\mathbf{k}} [G^\phi]_{ab}^{-1} \phi_{b\mathbf{k}}^\dagger = I_{4\times 4}, \tag{99}$$

where $[G^\phi]_{ab} = \phi_{a\mathbf{k}}^\dagger \phi_{b\mathbf{k}}$. And we have the equation

$$P_{\text{Bloch}} = I_{4\times 4} P_{\text{Bloch}} I_{4\times 4} = P_{\text{probe}} P_{\text{Bloch}} P_{\text{probe}} \tag{100}$$

$$= \left(\sum_{ab} \phi_{a\mathbf{k}} [G^\phi]_{ab}^{-1} \phi_{b\mathbf{k}}^\dagger\right) P_{\text{Bloch}} \left(\sum_{cd} \phi_{c\mathbf{k}} [G^\phi]_{cd}^{-1} \phi_{d\mathbf{k}}^\dagger\right). \tag{101}$$

And $\phi_b^\dagger P_{\text{Bloch}} \phi_c$ is the measured matrix element in ARPES $\langle W_b|\hat{P}|W_c\rangle$ in the four-band tight-binding basis. So the Bloch states projector can be expressed in terms of the measured matrix elements and the known projected probes $\phi_{a\mathbf{k}}$.

For the QGT, we have

$$Q_{ij}^{nn'}(\mathbf{k}) = u_{n\mathbf{k}}^\dagger \partial_i P_{\text{Bloch}} \partial_j P_{\text{Bloch}} u_{n'\mathbf{k}}; \tag{102}$$

$$g_{ij}^{nn'} = \frac{1}{2} u_{n\mathbf{k}}^\dagger \{\partial_i P_{\text{Bloch}}, \partial_j P_{\text{Bloch}}\} u_{n'\mathbf{k}}; \tag{103}$$

$$\Omega_{ij}^{nn'} = i u_{n\mathbf{k}}^\dagger [\partial_i P_{\text{Bloch}}, \partial_j P_{\text{Bloch}}] u_{n'\mathbf{k}}. \tag{104}$$

The nonzero part of the spectrum of the quantum metric $g_{ij}$ is identical to

$$\frac{1}{2}\{\partial_i P_{\text{Bloch}}, \partial_j P_{\text{Bloch}}\} P_{\text{Bloch}}, \tag{105}$$

and nonzero spectrum of the non-Abelian Berry curvature is identical to

$$i[\partial_i P_{\text{Bloch}}, \partial_j P_{\text{Bloch}}] P_{\text{Bloch}}. \tag{106}$$

## VIII. INS INTENSITY FOR PHONON BANDS

The coherent dynamical structure factor for all vibration modes are written as [4]

$$\mathcal{S}^{(i)}(\mathbf{q}, \omega) = \frac{(2\pi)^3}{v_0} \frac{|F^{(i)}(\mathbf{q})|^2}{2\omega_i(\mathbf{k})} \delta(\omega - \omega^{(i)}(\mathbf{k}));$$
$$F^{(i)}(\mathbf{q}) = \sum_d \frac{b_{d,\text{coh}}}{\sqrt{m_d}} \mathbf{q} \cdot \boldsymbol{\xi}_i^{(d)} e^{i\mathbf{G}\cdot\mathbf{r}_d} e^{-W_d} \tag{107}$$

where $\mathbf{G} = \mathbf{q} - \mathbf{k}$ is a reciprocal lattice vector, and $m_d, \mathbf{r}_d, b_{d,\text{coh}}, \boldsymbol{\xi}_d^{(i)}$ and $W_d$ denote the mass, position, coherent scattering length, the polarization vector and the Debye-Waller factor of the $d$th atom in the primitive unit cell.

One can show $|F^{(i)}(\mathbf{q})|^2$ is exactly the wavefunction form factor. The phonon's eigenvector $\boldsymbol{\xi}^{(i)}(\mathbf{k})$ is a $3n$-dimensional complex vector where $n$ is the number of atoms in the primitive unit cell. And the probe state is defined by

$$W_{\alpha d}(\mathbf{q}) = \frac{b_{d,\text{coh}}}{\sqrt{m_d}} q^\alpha e^{-i\mathbf{G}\cdot\mathbf{r}_d} e^{-W_d}. \tag{108}$$

And our conclusion in the non-degenerate electronic bands can be directly applied to non-degenerate phonon bands. Notice that the coherent phonon scattering cross section is independent of the neutron spin, so the degrees of freedom are not enough to construct the WFF for the degenerate phonon bands.

One may notice the phase factor in $F^{(i)}(\mathbf{q})$ is different from the textbook $e^{i\mathbf{q}\cdot\mathbf{r}_d}$. The reason is we adapt a different convention of Fourier transformation. To solve the phonon frequency and eigenvector, we need to solve the equation of motion

$$m_d \ddot{\mathbf{u}}_{ld}(t) = -\sum_{l'd'} \boldsymbol{\Phi}_{dd'}(l - l') \cdot \mathbf{u}_{l'd'}(t), \tag{109}$$

where $\boldsymbol{\Phi}_{dd'}(l - l')$ is the force-constant matrix, which is $3 \times 3$ and $\mathbf{u}_{ld}$ is the displacement of the atoms from the equilibrium positions. We can write an ansatz solution

$$\mathbf{u}_{ld}(t) = \frac{1}{\sqrt{m_d}} \boldsymbol{\xi}_d \exp(i\mathbf{k} \cdot \mathbf{R}_{ld} - i\omega t). \tag{110}$$

Then the equation of motion becomes

$$\omega^2 \boldsymbol{\xi}_d(\mathbf{k}) = \sum_{d'} \mathbf{D}_{dd'}(\mathbf{k}) \cdot \boldsymbol{\xi}_{d'}(\mathbf{k}), \tag{111}$$

where the $3n \times 3n$ dynamical matrix is given by

$$\mathbf{D}_{dd'}(\mathbf{k}) = \sum_{l'} \boldsymbol{\Phi}_{dd'}(-l') \exp(i\mathbf{k} \cdot (\mathbf{R}_{l'} + \mathbf{r}_{d'} - \mathbf{r}_d)). \tag{112}$$

One can check the dynamical matrix is Hermitian. The eigenvector of the dynamical matrix $\boldsymbol{\xi}^{(i)}(\mathbf{k})$ gives the phonon wavefunction. Thus, one can directly apply our discussion about Chern number in non-degenerate electronic band to phonon band system.

# IX. INS INTENSITY FOR MAGNON BANDS

In this section, we focus on the magnetic scattering [5, 6].

### 1. Unpolarized cross section

We start with the simplest unpolarized cross section, which is given by

$$\frac{d^2\sigma}{d\Omega dE'} \propto \sum_{ld}\sum_{l'd'} \exp\{i\mathbf{q}\cdot(\mathbf{R}_{l'd'}-\mathbf{R}_{ld})\} \frac{1}{2}g_d F_d^*(\mathbf{q}) \frac{1}{2}g_{d'} F_{d'}(\mathbf{q}) \times \frac{1}{2\pi}\int_{-\infty}^{\infty} dt \exp(-i\omega t) \left\langle \hat{\mathbf{S}}_{ld}^{(\perp)} \cdot \hat{\mathbf{S}}_{l'd'}^{(\perp)}(t) \right\rangle, \quad (113)$$

where $\hat{\mathbf{S}}^{(\perp)} = \hat{\mathbf{q}} \times (\hat{\mathbf{S}}^{(\perp)} \times \hat{\mathbf{q}})$ is the spin component perpendicular to the transfer momentum $\mathbf{q}$, $g_d, F_d(\mathbf{q})$ are the Landé g-factor, the magnetic form factor for the $d$th atom in the magnetic unit cell. If the system possesses $S_z$ conservation, the spin-spin correlation function can be expanded as

$$\left\langle \hat{\mathbf{S}}_{ld}^{(\perp)} \cdot \hat{\mathbf{S}}_{l'd'}^{(\perp)}(t) \right\rangle = (\delta_{\alpha\beta} - \hat{q}_\alpha \hat{q}_\beta) \langle \hat{S}_{ld}^\alpha \hat{S}_{l'd'}^\beta(t) \rangle. \quad (114)$$

Since $\langle \hat{S}^x \hat{S}^x \rangle = \langle \hat{S}^y \hat{S}^y \rangle = \frac{1}{4}(\langle \hat{S}^+ \hat{S}^- \rangle + \langle \hat{S}^- \hat{S}^+ \rangle)$, we have

$$\left\langle \hat{\mathbf{S}}_{ld}^{(\perp)} \cdot \hat{\mathbf{S}}_{l'd'}^{(\perp)}(t) \right\rangle = \frac{1}{2}(1+\hat{q}_z^2)(\langle \hat{S}_{ld}^+ \hat{S}_{l'd'}^-(t) \rangle + \langle \hat{S}_{ld}^- \hat{S}_{l'd'}^+(t) \rangle), \quad (115)$$

where we've ignored the $\langle \hat{S}_z \hat{S}_z \rangle$ term which won't contribute to the inelastic scattering process.

To see how the cross section is related to the WFF, let's analyze the ferromagnetic magnon with $U(1)$ symmetry. We perform the Holstein-Primakoff transformation

$$S^+ = \sqrt{2S}\hat{a}; \quad S^- = \sqrt{2S}\hat{a}^\dagger; \quad S^z = S - \hat{a}^\dagger \hat{a}, \quad (116)$$

and the magnon Bloch Hamiltonian is given by

$$\hat{H} = \sum_{d,d'} \hat{a}_d^\dagger(\mathbf{k}) H_{dd'}(\mathbf{k}) \hat{a}_{d'}(\mathbf{k}), \quad (117)$$

where $\hat{a}_d(\mathbf{k}) = \frac{1}{N}\sum_l e^{i\mathbf{k}\cdot\mathbf{R}_{ld}} \hat{a}_{ld}$. To find the eigenmode, we write the Heisenberg equation of motion, which is give by

$$i\frac{\partial \hat{a}_d(\mathbf{k},t)}{\partial t} = [\hat{a}_d(\mathbf{k}), \hat{H}](t) = \sum_{d'} H_{dd'}(\mathbf{k}) \hat{a}_{d'}(\mathbf{k},t). \quad (118)$$

By diagonalizing the dynamical matrix, which coincides with the Bloch Hamiltonian matrix for the ferro-magnon case, we get the eigenmodes $\hat{\alpha}_n(\mathbf{k})$:

$$\hat{\alpha}_n(\mathbf{k}) = \sum_d U_{dn}^*(\mathbf{k}) \hat{a}_d(\mathbf{k}), \quad (119)$$

where the unitary matrix diagonalizes the Bloch Hamiltonian: $U^\dagger H U = E$. We define the magnon wavefunction as the eigenvector of the dynamical matrix $\psi_d^{(n)}(\mathbf{k}) \equiv U_{dn}(\mathbf{k})$.

By using the eigenmodes operators $\hat{\alpha}_n(\mathbf{k}), \hat{\alpha}_n^\dagger(\mathbf{k})$, the cross section can be rewritten as

$$\begin{aligned}\frac{d^2\sigma}{d\Omega dE'} &\propto \sum_{dd'} \exp\{i\mathbf{G}\cdot(\mathbf{r}_{d'}-\mathbf{r}_d)\} \frac{1}{2}g_d F_d^*(\mathbf{q}) \frac{1}{2}g_{d'} F_{d'}(\mathbf{q}) \\ &\times \frac{1}{2}(1+\hat{q}_z^2) \sum_n [\psi_d^{(n)} \psi_{d'}^{(n)*}(n_B(\varepsilon_{n\mathbf{k}})+1)\delta(\omega-\varepsilon_{n\mathbf{k}}) + \psi_{d'}^{(n)} \psi_d^{(n)*} n_B(\varepsilon_{n\mathbf{k}})\delta(\omega+\varepsilon_{n\mathbf{k}})],\end{aligned} \quad (120)$$

and the two terms in the last row stand for absorbing and emitting a magnon and $n_B(\varepsilon)$ is the Bose-Einstein distribution function. Only the absorbing term is non-zero at zero temperature and without losing generality, we only consider term below.

By defining the components of the probe state wavefunction $W_d(\mathbf{q})$:

$$W_d(\mathbf{q}) = \frac{1}{2}(1+\hat{q}_z^2)e^{i\mathbf{G}\cdot\mathbf{r}_d}(\frac{1}{2}g_d F_d(\mathbf{q})), \tag{121}$$

the cross section can be rewritten as

$$\frac{\mathrm{d}^2\sigma}{\mathrm{d}\Omega\mathrm{d}E'} \propto \sum_n |\langle\psi_n(\mathbf{k})|W(\mathbf{q})\rangle|^2 \delta(\omega-\varepsilon_{n\mathbf{k}})(n_B(\omega)+1). \tag{122}$$

It bears the same structure as the spin-polarized spectral function for the non-degenerate electronic band case. So when the ferro-magnon band features Weyl points, we predict spectral nodal arc connecting Weyl points with opposite charges.

### 2. Polarized cross section

Now we consider the case where the incident neutrons has a non-zero polarization $\mathbf{P}$. The cross section now is

$$\begin{aligned}\frac{\mathrm{d}^2\sigma}{\mathrm{d}\Omega\mathrm{d}E'} &\propto \sum_{l,\mathbf{d}}\sum_{l'\mathbf{d}'} \exp\{i\mathbf{q}\cdot(\mathbf{R}_{l'd'}-\mathbf{R}_{ld})\}\frac{1}{2}g_d F_d^*(\mathbf{q})\frac{1}{2}g_{d'} F_{d'}(\mathbf{q}) \\ &\times \frac{1}{2\pi\hbar}\int_{-\infty}^{\infty}\mathrm{d}t\,\exp(-i\omega t)\left\{\left\langle\hat{\mathbf{S}}_{ld}^{(\perp)}\cdot\hat{\mathbf{S}}_{l'd'}^{(\perp)}(t)\right\rangle + i\mathbf{P}\cdot\left\langle\hat{\mathbf{S}}_{ld}^{(\perp)}\times\hat{\mathbf{S}}_{l'd'}^{(\perp)}(t)\right\rangle\right\}.\end{aligned} \tag{123}$$

For the $S_z$ conservation system, the cross section can be expanded as

$$\begin{aligned}&\left\langle\hat{\mathbf{S}}_{ld}^{(\perp)}\cdot\hat{\mathbf{S}}_{l'd'}^{(\perp)}(t)\right\rangle + i\mathbf{P}\cdot\left\langle\hat{\mathbf{S}}_{ld}^{(\perp)}\times\hat{\mathbf{S}}_{l'd'}^{(\perp)}(t)\right\rangle \\ =&\frac{1}{4}\left\{(1+(\hat{\mathbf{q}}\cdot\boldsymbol{\eta})^2+2(\mathbf{P}\cdot\hat{\mathbf{q}})(\hat{\mathbf{q}}\cdot\boldsymbol{\eta}))\langle\hat{S}_{ld}^-\hat{S}_{l'd'}^+(t)\rangle + (1+(\hat{\mathbf{q}}\cdot\boldsymbol{\eta})^2-2(\mathbf{P}\cdot\hat{\mathbf{q}})(\hat{\mathbf{q}}\cdot\boldsymbol{\eta}))\langle\hat{S}_{ld}^+\hat{S}_{l'd'}^-(t)\rangle\right\},\end{aligned} \tag{124}$$

where $\boldsymbol{\eta}$ is the quantization axis (which we choose as the $z$-axis).

The polarized INS can be used to extract the WFF in antiferromagnetic system. The Holstein-Primakoff transformation now becomes

$$\hat{S}^+ = \sqrt{2S}\hat{a}, \quad \hat{S}^z = S - \hat{a}^\dagger\hat{a}; \tag{125}$$

for the up spins and

$$\hat{S}^+ = -\sqrt{2S}\hat{b}^\dagger, \quad \hat{S}^z = \hat{b}^\dagger\hat{b} - S. \tag{126}$$

By choosing the basis $\hat{\Psi}(\mathbf{k}) = (\hat{a}_{i=1,\ldots,n}(\mathbf{k}), \hat{b}^\dagger_{i=1,\ldots,n}(-\mathbf{k}))$, the Bloch Hamiltonian is given by

$$\hat{H} = \sum_{\mathbf{k}} \hat{\Psi}^\dagger(\mathbf{k}) H(\mathbf{k}) \hat{\Psi}(\mathbf{k}), \tag{127}$$

which has the same structure as the Bogoliubov-de-Gennes (BdG) Hamiltonian. To find the eigenmodes, we need to solve the eigensystem of the non-Hermitian dynamical matrix $D(\mathbf{k})$:

$$D(\mathbf{k}) = \begin{pmatrix} I_{n\times n} & \\ & -I_{n\times n} \end{pmatrix} H(\mathbf{k}) = \sigma_z H(\mathbf{k}). \tag{128}$$

Suppose we have $U^{-1}(\mathbf{k})D(\mathbf{k})U(\mathbf{k}) = E(\mathbf{k})$, one can check $U^\dagger(\mathbf{k})\sigma_z U(\mathbf{k}) = \sigma_z$. In other words, the similarity transformation $U(\mathbf{k})$ preserves the canonical commutation relation.

Denoting the eigenmode basis as $\hat{\Phi}(\mathbf{k}) = (\hat{\alpha}_{m=1,\ldots,n}(\mathbf{k}), \hat{\beta}^\dagger_{m=1,\ldots,n}(\mathbf{k}))$ and we have $\hat{\Phi}(\mathbf{k}) = U(\mathbf{k})\hat{\Psi}(\mathbf{k})$. Substitute it back to the original cross section, and for the 1-magnon absorbing term we have

$$\frac{d^2\sigma}{d\Omega dE'} \propto \sum_{d\mathbf{d}'} \exp\{i\mathbf{G}\cdot(\mathbf{r}_{d'} - \mathbf{r}_d)\} \frac{1}{2}g_d F_d^*(\mathbf{q})\frac{1}{2}g_{d'}F_{d'}(\mathbf{q})$$

$$\times \frac{1}{4}\{(1 + (\hat{\mathbf{q}}\cdot\boldsymbol{\eta})^2 + 2(\mathbf{P}\cdot\hat{\mathbf{q}})(\hat{\mathbf{q}}\cdot\boldsymbol{\eta}))\sum_{m=n+1}^{2n} \psi_{d'}^{(m)*}(\mathbf{k})\psi_d^{(m)}(\mathbf{k})\delta(\omega - \varepsilon_{n\mathbf{k}})$$

$$+(1 + (\hat{\mathbf{q}}\cdot\boldsymbol{\eta})^2 - 2(\mathbf{P}\cdot\hat{\mathbf{q}}))\sum_{m=1}^{n} \psi_{d'}^{(m)}(\mathbf{k})\psi_d^{(m)*}(\mathbf{k})\delta(\omega - \varepsilon_{n\mathbf{k}})\} \quad (129)$$

$$=(1 + (\hat{\mathbf{q}}\cdot\boldsymbol{\eta})^2 - 2(\mathbf{P}\cdot\hat{\mathbf{q}}))\sum_{m=1}^{n} |\langle\psi_m(\mathbf{k})|W_1(\mathbf{q})\rangle|^2\delta(\omega - \varepsilon_{m\mathbf{k}})$$

$$+(1 + (\hat{\mathbf{q}}\cdot\boldsymbol{\eta})^2 + 2(\mathbf{P}\cdot\hat{\mathbf{q}}))\sum_{m=n+1}^{2n} |\langle\psi_m(\mathbf{k})|W_2(\mathbf{q})\rangle|^2\delta(\omega - \varepsilon_{m\mathbf{k}})$$

where $\psi_d^{(m)}(\mathbf{k}) = U_{dm}(\mathbf{k})$, $m = 1,\ldots,n$ with $S_z$ quantum number $+1$ and $\psi_d^{(m)}(\mathbf{k}) = U_{dm}(\mathbf{k})$, $m = n+1,\ldots,2n$ with $S_z$ quantum number $-1$. The probe state is defined by

$$W_{1,d}(\mathbf{q}) = (\frac{1}{2}g_d F_d(\mathbf{q}))e^{i\mathbf{G}\cdot\mathbf{r}_d} \quad (130)$$

and $W_{2,d} = W_{1,d}^*$. For antiferromagnetic system with $PT$ symmetry, there can be degeneracies between bands with opposite $S_z$ quantum number. In this case, the zero of $|\langle\psi_m(\mathbf{k})|W_1(\mathbf{q})\rangle|^2$ and $|\langle\psi_{m+n}(\mathbf{k})|W_2(\mathbf{q})\rangle|^2$ will not coincide. Now we consider the INS intensity with two different polarizations:

$$\begin{aligned}I(\mathbf{q},\omega,\mathbf{P}_1) =&[(1 + (\hat{\mathbf{q}}\cdot\boldsymbol{\eta})^2 - 2(\mathbf{P}_1\cdot\hat{\mathbf{q}}))|\langle\psi_m(\mathbf{k})|W_1(\mathbf{q})\rangle|^2 \\ &+(1 + (\hat{\mathbf{q}}\cdot\boldsymbol{\eta})^2 + 2(\mathbf{P}_1\cdot\hat{\mathbf{q}}))|\langle\psi_{m+n}(\mathbf{k})|W_2(\mathbf{q})\rangle|^2]\delta(\omega - \varepsilon_{m\mathbf{k}}); \\ I(\mathbf{q},\omega,\mathbf{P}_2) =&[(1 + (\hat{\mathbf{q}}\cdot\boldsymbol{\eta})^2 - 2(\mathbf{P}_2\cdot\hat{\mathbf{q}}))|\langle\psi_m(\mathbf{k})|W_1(\mathbf{q})\rangle|^2 \\ &+(1 + (\hat{\mathbf{q}}\cdot\boldsymbol{\eta})^2 + 2(\mathbf{P}_2\cdot\hat{\mathbf{q}}))|\langle\psi_{m+n}(\mathbf{k})|W_2(\mathbf{q})\rangle|^2]\delta(\omega - \varepsilon_{m\mathbf{k}})\end{aligned} \quad (131)$$

Then one can solve

$$|\langle\psi_m(\mathbf{k})|W_1(\mathbf{q})\rangle|^2 = \frac{-(1 + (\hat{\mathbf{q}}\cdot\boldsymbol{\eta})^2 + 2(\mathbf{P}_2\cdot\hat{\mathbf{q}}))I(\mathbf{P}_1) + (1 + (\hat{\mathbf{q}}\cdot\boldsymbol{\eta})^2 + 2(\mathbf{P}_1\cdot\hat{\mathbf{q}}))I(\mathbf{P}_2)}{4(1 + (\hat{\mathbf{q}}\cdot\boldsymbol{\eta})^2)(\mathbf{P}_1 - \mathbf{P}_2)\cdot\hat{\mathbf{q}}}$$

$$|\langle\psi_{m+n}(\mathbf{k})|W_2(\mathbf{q})\rangle|^2 = \frac{(1 + (\hat{\mathbf{q}}\cdot\boldsymbol{\eta})^2 - 2(\mathbf{P}_2\cdot\hat{\mathbf{q}}))I(\mathbf{P}_1) - (1 + (\hat{\mathbf{q}}\cdot\boldsymbol{\eta})^2 - 2(\mathbf{P}_1\cdot\hat{\mathbf{q}}))I(\mathbf{P}_2)}{4(1 + (\hat{\mathbf{q}}\cdot\boldsymbol{\eta})^2)(\mathbf{P}_1 - \mathbf{P}_2)\cdot\hat{\mathbf{q}}}. \quad (132)$$

### 3. Full polarization analysis

We can extract more information from the full polarization analysis. The polarization of the outgoing neutron beam is given by

$$\mathbf{P}(\frac{d^2\sigma}{d\Omega dE'}) \propto \sum_{l,\mathbf{d}}\sum_{l'\mathbf{d}'} \exp\{i\mathbf{q}\cdot(\mathbf{R}_{l'd'} - \mathbf{R}_{ld})\}\frac{1}{2}g_d F_d^*(\mathbf{q})\frac{1}{2}g_{d'}F_{d'}(\mathbf{q})$$

$$\times \int_{-\infty}^{\infty}\frac{dt}{2\pi}e^{-i\omega t}\{\hat{\mathbf{S}}_{ld}^{(\perp)}(\mathbf{P}\cdot\hat{\mathbf{S}}_{l'd'}^{(\perp)}(t)) + (\mathbf{P}\cdot\hat{\mathbf{S}}_{ld}^{(\perp)})\hat{\mathbf{S}}_{l'd'}^{(\perp)}(t) - \mathbf{P}(\hat{\mathbf{S}}_{ld}^{(\perp)}\cdot\hat{\mathbf{S}}_{l'd'}^{(\perp)}(t)) - i(\hat{\mathbf{S}}_{ld}^{(\perp)}\times\hat{\mathbf{S}}_{l'd'}^{(\perp)}(t))\}. \quad (133)$$

Now we define an intensity vector $\mathcal{I}$

$$\mathcal{I}(\mathbf{q},\mathbf{P},\omega) = (\frac{d^2\sigma}{d\Omega dE'}, \mathbf{P}(\frac{d^2\sigma}{d\Omega dE'})). \quad (134)$$

It is straightforward to show that $(\mathcal{I}^0 \pm \mathcal{I}^j)/2$ measures the cross section of the outgoing neutron beaming whose polarization is along the $\pm i$-axis.

Now we expand the spin-spin correlation function in the intensity vector $\mathcal{I}(\mathbf{q}, \mathbf{P}, \omega)$ as:

$$\sum_{\alpha\beta\gamma} C^\alpha_{\beta\gamma}(\hat{\mathbf{q}}, \mathbf{P}) \langle \hat{S}^\beta_{ld} \hat{S}^\gamma_{l'd'}(t) \rangle. \tag{135}$$

We call the rank-3 tensor $C^\alpha_{\beta\gamma}$ as the kinematic tensor. By definition, we have

$$\begin{aligned}
C^0_{\beta\gamma} &= \delta_{\beta\gamma} - \hat{\mathbf{q}}_\beta \hat{\mathbf{q}}_\gamma + i[\mathbf{P}_\alpha \epsilon^\alpha_{\beta\gamma} + (\mathbf{P}\times\hat{\mathbf{q}})_\beta \hat{\mathbf{q}}_\gamma - (\mathbf{P}\times\hat{\mathbf{q}})_\gamma \hat{\mathbf{q}}_\beta]; \\
C^\alpha_{\beta\gamma} &= (\delta^\alpha_\beta - \hat{\mathbf{q}}^\alpha \hat{\mathbf{q}}_\beta)(\mathbf{P}_\gamma - (\mathbf{P}\cdot\mathbf{q})\mathbf{q}_\gamma) + (\beta \longleftrightarrow \gamma) \\
&\quad + \mathbf{P}^\alpha(\delta_{\beta\gamma} - \hat{\mathbf{q}}_\beta \hat{\mathbf{q}}_\gamma) - i(\epsilon^\alpha_{\beta\gamma} - \epsilon^\alpha_{\beta\gamma'}\hat{\mathbf{q}}^{\gamma'}\hat{\mathbf{q}}_\gamma + \epsilon^\alpha_{\gamma\gamma'}\hat{\mathbf{q}}^{\gamma'}\hat{\mathbf{q}}_\beta).
\end{aligned} \tag{136}$$

The tensor can be further simplified by noticing that

$$C^\alpha_{\beta\gamma} \hat{\mathbf{q}}_\gamma = 0. \tag{137}$$

By doing a basis transformation, we can reduce the kinematic tensor. Now we choose a new basis in the real space

$$R = \begin{pmatrix} \hat{\mathbf{e}}_1 & \hat{\mathbf{e}}_2 & \hat{\mathbf{e}}_3 \end{pmatrix}, \tag{138}$$

and $\hat{\mathbf{e}}_3 = \hat{\mathbf{q}}$. The non-zero bloch of the transformed kinematic tensor, $\tilde{C}^\alpha = R^T C^\alpha R$, is given by

$$\begin{aligned}
\tilde{C}^0_{ab}(\hat{\mathbf{q}}, \mathbf{P}) &= \begin{pmatrix} 1 & i\mathbf{P}\cdot\hat{\mathbf{q}} \\ -i\mathbf{P}\cdot\hat{\mathbf{q}} & 1 \end{pmatrix} \\
\tilde{C}^\alpha_{ab}(\hat{\mathbf{q}}, \mathbf{P}) &= (\hat{\mathbf{e}}_a)_\alpha (\mathbf{P}\cdot\hat{\mathbf{e}}_b) + (\mathbf{P}\cdot\hat{\mathbf{e}}_a)(\hat{\mathbf{e}}_b)_\alpha - \delta_{ab}\mathbf{P}_\alpha - i\epsilon_{ab3}\mathbf{q}_\alpha.
\end{aligned} \tag{139}$$

This will be the starting point of the full polarization analysis.

One may question that the kinetic matrices defined above do not necessarily have positive eigenvalues so is it legal to call them "density matrices". In realistic experiments, what we measure is the scattering cross section of outgoing neutron polarized along one specific direction, let's say, $+x$. And the corresponding spin-spin correlation function in the intensity is given by

$$\frac{1}{2}\sum_{\beta\gamma}(C^0 + C^x)_{\beta\gamma} \langle \hat{S}^\beta_{ld} \hat{S}^\gamma_{l'd'}(t) \rangle, \tag{140}$$

and now the Hermitian matrix $\frac{1}{2}(C^0 + C^x)$ have positive eigenvalues (see an example at the end of this section) and one should view $\frac{1}{2}(C^0 + C^x)$ as the true density matrix. Actually, for a single measurement the associated "density matrix" must be positive definite to get a physical intensity measurement (which is guaranteed to be real, positive).

We consider a general non-collinear magnetic system. We do a local rotation to the spins so in the ground state the rotated spins $\hat{\mathbf{S}}'_{ld}$ are in the ferromagnetic order:

$$\hat{\mathbf{S}}_{ld} = R_d \hat{\mathbf{S}}'_{ld}, \tag{141}$$

and we perform the Holstein-Primakoff transformation to the rotated spins. And we have

$$\hat{\mathbf{S}}'_{ld} = \sqrt{\frac{S_j}{2}}(\bar{\mathbf{u}}_d \hat{a}_{ld} + \mathbf{u}_d \hat{a}^\dagger_{ld}) + \mathbf{v}_d(S_j - \hat{a}^\dagger_{ld}\hat{a}_{ld}), \tag{142}$$

where $u^\alpha_d = R^{\alpha 1}_d + iR^{\alpha 2}_d, v^\alpha_d = R^{\alpha 3}_d$. The intensity vector can be written as

$$\mathcal{I}^\alpha(\mathbf{q}, \mathbf{P}, \omega) = \int_{-\infty}^\infty \frac{dt}{2\pi} e^{-i\omega t} C^\alpha_{\beta\gamma}(\hat{\mathbf{q}}, \mathbf{P}) \langle \hat{\Psi}^\dagger(\mathbf{k}) W^\beta(\mathbf{q}) W^{\gamma*}(\mathbf{q}) \hat{\Psi}(\mathbf{k}, t) \rangle. \tag{143}$$

The basis $\hat{\Psi}(\mathbf{k})$ is defined as

$$\hat{\Psi}(\mathbf{k}) = (\hat{a}_1(\mathbf{k}), ..., \hat{a}_n(\mathbf{k}); \hat{a}^\dagger_1(-\mathbf{k}), ..., \hat{a}^\dagger_n(-\mathbf{k}))^T \tag{144}$$

and the spectroscopic probe state $W^\alpha(\mathbf{q})$

$$W^\alpha(\mathbf{q}) = (\frac{1}{2}g_1 F_1^* e^{i\mathbf{G}\cdot\mathbf{r}_1} u^\alpha_1, ... \frac{1}{2}g_n F_n^* e^{i\mathbf{G}\cdot\mathbf{r}_n} u^\alpha_n; \frac{1}{2}g_1 F_1 e^{-i\mathbf{G}\cdot\mathbf{r}_1} \bar{u}^\alpha_1, ..., \frac{1}{2}g_n F_n e^{-i\mathbf{G}\cdot\mathbf{r}_n} \bar{u}^\alpha_n)^T. \tag{145}$$

Here $u_d^\alpha = R_d^{\alpha 1} + i R_d^{\alpha 2}$. By reducing the kinematic tensor, we have

$$\mathcal{I}^\alpha(\mathbf{q}, \mathbf{P}, \omega) = \int_{-\infty}^{\infty} \frac{dt}{2\pi} e^{-i\omega t} \tilde{C}_{ab}^\alpha(\hat{\mathbf{q}}, \mathbf{P}) \langle \hat{\Psi}^\dagger(\mathbf{k}) \tilde{W}_a(\mathbf{q}) \tilde{W}_b^*(\mathbf{q}) \hat{\Psi}(\mathbf{k}, t) \rangle, \tag{146}$$

where $\tilde{W}_a = \hat{\mathbf{e}}_a^\beta W^\beta$.

To find the eigenmodes, we need to solve the eigensystem of the dynamical matrix

$$D(\mathbf{k}) = \begin{pmatrix} I_{n \times n} & \\ & -I_{n \times n} \end{pmatrix} H(\mathbf{k}), \tag{147}$$

where $H(\mathbf{k})$ is the Bloch-BdG Hamiltonian. This is similar to what we've done in the antiferro-magnon. The final intensity vector is given by

$$\mathcal{I}^\alpha(\mathbf{q}, \omega) = \sum_{ab} C_{ab}^\alpha(\hat{\mathbf{q}}, \mathbf{P}) \sum_m \sum_{dd'} \psi_d^{(m)*}(\mathbf{k}) \tilde{W}_d^a \tilde{W}_{d'}^{b,*} \psi_{d'}^{(m)}(\mathbf{k}) \delta(\omega - g_{mm}\varepsilon_m(\mathbf{k}))(n_B(\omega) + \frac{1}{2}(1 + g_{mm})). \tag{148}$$

Now we focus on the degenerate doublet on a specific energy branch. Its contribution to the intensity vector is given by

$$\mathcal{I}^\alpha(\mathbf{q}, \mathbf{P}, \omega) = \sum_{ab} C_{ab}^\alpha(\hat{\mathbf{q}}, \mathbf{P}) \langle \tilde{W}_b | \hat{P}(\mathbf{k}) | \tilde{W}_a \rangle \delta(\omega - \varepsilon_n(\mathbf{k})), \tag{149}$$

where $\hat{P}(\mathbf{k}) = |\psi_1(\mathbf{k})\rangle \langle \psi_1(\mathbf{k})| + |\psi_2(\mathbf{k})\rangle \langle \psi_2(\mathbf{k})|$. Now our goal is to solve the determinant (which is the WFF determinant in the degenerate band case)

$$\det[\langle \tilde{W}_b | \hat{P}(\mathbf{k}) | \tilde{W}_a \rangle]. \tag{150}$$

Now that Eq. ((149)) is actually an linear system and the matrix elements, $\langle \tilde{W}_b | \hat{P}(\mathbf{k}) | \tilde{W}_a \rangle$, are the unknown variables. The linear system can be solved by the expansion

$$C^\alpha = \frac{1}{2} \sum_{j=0}^{4} \sigma^\beta \text{Tr}[\sigma^\beta C^\alpha] = \sum_{j=0}^{4} \sigma_\beta B_{\beta\alpha}, \tag{151}$$

and

$$\begin{aligned} \{B_{0\alpha}\} &= (1, -\hat{\mathbf{q}}(\hat{\mathbf{q}} \cdot \mathbf{P})) \\ \{B_{1\alpha}\} &= (0, (\mathbf{P} \cdot \hat{\mathbf{e}}_2)\hat{\mathbf{e}}_1 + (\mathbf{P} \cdot \hat{\mathbf{e}}_1)\hat{\mathbf{e}}_2) \\ \{B_{2\alpha}\} &= (-\hat{\mathbf{q}} \cdot \mathbf{P}, \hat{\mathbf{q}}) \\ \{B_{3\alpha}\} &= (0, (\mathbf{P} \cdot \hat{\mathbf{e}}_1)\hat{\mathbf{e}}_1 - (\mathbf{P} \cdot \hat{\mathbf{e}}_2)\hat{\mathbf{e}}_2). \end{aligned} \tag{152}$$

Then the intensity vector can be rewritten as

$$\begin{aligned} \mathcal{I}^\alpha(\mathbf{q}, \mathbf{P}, \omega) &= \sum_{ab} \tilde{C}_{ab}^\alpha(\hat{\mathbf{q}}, \mathbf{P}) \langle \tilde{W}_b | \hat{P}(\mathbf{k}) | \tilde{W}_a \rangle \delta(\omega - \varepsilon_m(\mathbf{k})) \\ &= \frac{1}{2} \text{Tr}[\sigma_\beta F] B_{\beta\alpha} \delta(\omega - \varepsilon_m(\mathbf{k})). \end{aligned} \tag{153}$$

It is straightforward to show

$$\det F = \frac{\text{Tr}[\sigma_0 F]^2 - \sum_{i=1}^{3} \text{Tr}[\sigma_i F]^2}{4}, \tag{154}$$

and

$$\frac{1}{2} \text{Tr}[\sigma_\beta F] = \sum_\alpha \mathcal{I}^\alpha(\mathbf{q}, \mathbf{P}, \omega) B_{\alpha\beta}^{-1} / \delta(\omega - \varepsilon_m(\mathbf{k})) \equiv \tilde{\mathcal{I}}^\beta(\mathbf{q}, \mathbf{P}, \omega) / \delta(\omega - \varepsilon_m(\mathbf{k})). \tag{155}$$

So we have successfully express the WFF, $\det[F]$, as the function of the rotated intensity vector $\tilde{\mathcal{I}}(\mathbf{q}, \mathbf{P}, \omega)$. And

$$\begin{aligned}
\{B_{\alpha 0}^{-1}\} &= \frac{1}{1-(\hat{\mathbf{q}}\cdot\mathbf{P})^2}(1, -\hat{\mathbf{q}}(\hat{\mathbf{q}}\cdot\mathbf{P})); \\
\{B_{\alpha 1}^{-1}\} &= \frac{1}{|\mathbf{P}^{(\perp)}|^2}(0, (\mathbf{P}\cdot\hat{\mathbf{e}}_2)\hat{\mathbf{e}}_1 - (\mathbf{P}\cdot\hat{\mathbf{e}}_1)\hat{\mathbf{e}}_2); \\
\{B_{\alpha 2}^{-1}\} &= \frac{1}{1-(\hat{\mathbf{q}}\cdot\mathbf{P})^2}(\hat{\mathbf{q}}\cdot\mathbf{P}, \hat{\mathbf{q}}); \\
\{B_{\alpha 3}^{-1}\} &= \frac{1}{|\mathbf{P}^{(\perp)}|^2}(0, (\mathbf{P}\cdot\hat{\mathbf{e}}_2)\hat{\mathbf{e}}_1 + (\mathbf{P}\cdot\hat{\mathbf{e}}_1)\hat{\mathbf{e}}_2).
\end{aligned} \tag{156}$$

Based on this, the final WFF determinant is given by

$$\det[F] = \left[\frac{\mathcal{I}_0^2 - (\mathcal{I}\cdot\hat{\mathbf{q}})^2}{1-(\mathbf{P}\cdot\hat{\mathbf{q}})^2} - \frac{|\mathcal{I} - (\mathcal{I}\cdot\hat{\mathbf{q}})\hat{\mathbf{q}}|^2}{|\mathbf{P} - (\mathbf{P}\cdot\hat{\mathbf{q}})\hat{\mathbf{q}}|^2}\right]. \tag{157}$$

If we assume the incident neutron is fully polarized $|\mathbf{P}| = 1$, we have

$$\det[F] = \frac{\mathcal{I}_0^2 - \sum_{j=1}^3 \mathcal{I}_i^2}{1-(\mathbf{P}\cdot\hat{\mathbf{q}})^2}. \tag{158}$$

Suppose $\hat{\mathbf{q}} = \mathbf{e}_3$ (this can always be done by rotating the lab frame) and $\hat{\mathbf{P}} = \mathbf{e}_2$, the kinematic matrices will take a simple form

$$C^0 = I_{2\times 2}, \; C^1 = \sigma_x, \; C^2 = -\sigma_z, \; C^3 = \sigma_y. \tag{159}$$

## X. ROLE OF SYMMETRY

### 1. Electronic band

We can consider the point group symmetries. For instance, we consider the multi-Weyl points protected by the $C_{4/6}$-rotation symmetries. Due to the rotation symmetries, the multi-Weyl points are fixed on the rotation axis. The topological selection rule dictates the existence of the spectral nodes, but the symmetry will fix the location of the nodes. We illustrate this point by a two-band $k \cdot p$ model, which is given by

$$H_{\text{eff}}^{(c)}(\mathbf{k}) = a k_+^n \sigma_+ + a^* k_-^n \sigma_- - v_z q_z \sigma_z, \tag{160}$$

and the basis $|\psi_\pm\rangle$, which are eigenstates of $C_m$ with eigenvalue $\alpha_{p/q}$, where

$$\alpha_p = \exp(i2\pi p/m + iF\pi/m), \tag{161}$$

and $p = 0, ..., m-1$. We consider the spinful case, where $F = 1$. For the symmetry protected band crossing, we have $p \neq q$. Without loss of generality, we consider the $C_4$ group and consider the case $p - q \mod m = 1$: $\alpha_p = e^{i\pi/4}, \alpha_q = e^{-i\pi/4}$. At the north (south) pole of a sphere enclosing the Weyl point, the eigenstates will have well-defined rotation quantum number. We consider the the eigenstate at the upper branch and

$$\hat{C}_4 |\psi_{\mathbf{k}_{n/s}}\rangle = e^{\mp i\pi/4} |\psi_{\mathbf{k}_{n/s}}\rangle. \tag{162}$$

Recall that the momentum of the free electron can be decomposed as $\mathbf{q} = \mathbf{k} + \mathbf{G}$. We choose $\mathbf{G}$ to be at the rotation axis such that

$$C^4(\mathbf{k}_{n/s} + \mathbf{G}) = \mathbf{k}_{n/s} + \mathbf{G}, \tag{163}$$

and

$$\hat{C}_4 |\mathbf{q}\sigma\rangle = e^{i\sigma\pi/4} |\mathbf{q}\sigma\rangle. \tag{164}$$

So the inner product, $\langle \psi_\mathbf{k} | \mathbf{q}\sigma \rangle$ will have zeros at the north pole of the sphere for $\sigma = +1$ and at the south pole for $\sigma = -1$.

### 2. Phonon band

We first manifest the consequence of the $PT$-symmetry. The inversion symmetry enforces

$$\Phi_{dd'}^{\alpha\beta}(l-l') = \Phi_{\bar{d}\bar{d}'}^{\alpha\beta}(\bar{l}-\bar{l}'), \tag{165}$$

where $\bar{l}\bar{d}$ satisfies $\mathbf{R}_{ld} = -\mathbf{R}_{\bar{l}\bar{d}}$. Recall the definition of the dynamical matrix $D^{\alpha\beta}(\mathbf{k})$:

$$\mathbf{D}_{dd'}(\mathbf{k}) = \frac{1}{\sqrt{m_d m_{d'}}} \sum_{l'} \mathbf{\Phi}_{dd'}(-l') \exp(i\mathbf{k}\cdot(\mathbf{R}_{l'}+\mathbf{r}_{d'}-\mathbf{r}_d)). \tag{166}$$

The inversion symmetry ensures $\mathbf{D}_{dd'}(\mathbf{k}) = (\mathbf{D}_{\bar{d}\bar{d}'}(-\mathbf{k}) = U\mathbf{D}(-\mathbf{k})U^T)_{dd'}$. Here $U$ is an orthogonal matrix denoting the permutation of the sublattices. And the time-reversal symmetry ensures $\mathbf{D}^*(\mathbf{k}) = \mathbf{D}(-\mathbf{k})$. So the $PT$-symmetry ensures this property

$$(U\mathcal{K})D(\mathbf{k})(U\mathcal{K})^{-1} = D(\mathbf{k}). \tag{167}$$

Suppose there are $2n$ atoms in the unit cell, and we can organize the basis such that $\bar{d} = n+d, n=1,...,n$. And the permutation matrix will become

$$U = \tau_x \otimes I_{n\times n}. \tag{168}$$

By a basis transformation we have $\tilde{D}(\mathbf{k}) = V^\dagger D(\mathbf{k})V$, and under $PT$-operation

$$(PT)\tilde{D}(\mathbf{k})(PT)^{-1} = (V^\dagger UV^*)\tilde{D}^*(\mathbf{k})(V^\dagger UV^*)^\dagger. \tag{169}$$

By choosing

$$V = \frac{1}{\sqrt{2}}\begin{pmatrix} 1 & -i \\ 1 & i \end{pmatrix} \otimes I_{n\times n}, \tag{170}$$

we have $V^\dagger UV^* = I_{2n\times 2n}$ and $\tilde{D}(\mathbf{k}) = \tilde{D}^*(\mathbf{k})$. In other words, in this new basis the Bloch Hamiltonian is real and its eigenvector $\tilde{\boldsymbol{\xi}}(\mathbf{k})$ can also chosen to be real. The probe state in this new basis is

$$\tilde{\phi}_{\alpha d}(\mathbf{q}) = (V^\dagger)_{dd'}\phi_{\alpha d'}(\mathbf{q}). \tag{171}$$

Notice that $V^\dagger = \frac{1}{\sqrt{2}}\begin{pmatrix} 1 & 1 \\ i & -i \end{pmatrix}$, and $\phi_{\alpha\bar{d}}(\mathbf{q}) = \phi_{\alpha d}^*(\mathbf{q})$, the probe state $\tilde{\phi}_{\alpha d}(\mathbf{q})$ is also a real vector. Thus our discussion about the Berry phase in non-degenerate electronic band can be directly applied to phonon band system.

### 3. Magnon band

Now we prove in the presence of $PT$-symmetry with $(PT)^2 = 1$, the Bloch BdG Hamiltonian can also be chosen as real. We first do the Holstein-Primakoff transformation to locally rotated spin operators

$$S'^+ = \sqrt{2S}\hat{a}, \quad S'^z = S - \hat{a}^\dagger \hat{a}, \tag{172}$$

and the $PT$-transformation to the spin operators

$$(PT)\mathbf{S}_{ld}(PT)^{-1} = -\mathbf{S}_{\bar{l}\bar{d}}, \tag{173}$$

we know how the $PT$-operation acts on the rotated spin operators:

$$\begin{aligned}(PT)\mathbf{S}'_{ld}(PT)^{-1} &= (PT)R_d^{-1}\mathbf{S}_{ld}(PT)^{-1} \\ &= -R_d^{-1}\mathbf{S}_{\bar{l}\bar{d}} \\ &= -R_d^{-1}R_{\bar{d}}\mathbf{S}'_{\bar{l}\bar{d}}.\end{aligned} \tag{174}$$

Notice that for the rotated spin operators, we have

$$\langle \mathbf{S}'_{ld} \rangle = \langle \mathbf{S}'_{\bar{l}\bar{d}} \rangle, \tag{175}$$

and the expectation value can be understood either as to the ground state or the thermal average. Notice that both the ground state and the thermal state are invariant under the $PT$-operation. So by taking the expectation value of both sides of Eq. ((174)), we have

$$R_d = R_{\bar{d}} R_x, \tag{176}$$

where $R_x$ is the proper rotation along the $x-$axis by $\pi$ angle, and we have

$$(PT)\mathbf{S}'_{ld}(PT)^{-1} = -R_x \mathbf{S}'_{\bar{l}\bar{d}}. \tag{177}$$

Based on this, we can find how the $PT$-operation acts on the magnon operators:

$$\begin{aligned}(PT)\hat{a}_{ld}(PT)^{-1} &= -\hat{a}_{l'd'}; \\ (PT)\hat{a}_d(\mathbf{k})(PT)^{-1} &= -\hat{a}_{d'}(\mathbf{k})\end{aligned} \tag{178}$$

where $\hat{a}_d(\mathbf{k}) = \frac{1}{\sqrt{N}} \sum_l e^{-i\mathbf{k}\cdot(\mathbf{R_l}+\mathbf{r_d})} \hat{a}_{ld}$. By defining the basis vector $\hat{\Psi}(\mathbf{k}) = (\hat{a}_1(\mathbf{k}), \hat{a}_2(\mathbf{k})...\hat{a}_n(\mathbf{k}); \hat{a}_1^\dagger(-\mathbf{k}), ..., \hat{a}_n(-\mathbf{k}))$, the $PT$-operation in this basis acts as

$$PT = \text{Permu.} \times \mathcal{K}. \tag{179}$$

By rearrangement of the sublattice indices, we can have

$$\text{Permu.} = \tau_x \otimes I_{\frac{n}{2}\times \frac{n}{2}}. \tag{180}$$

Like we've done in the phonon band case, by defining a new basis $\tilde{\hat{\Psi}}(\mathbf{k}) = V\hat{\Psi}(\mathbf{k})$, the $PT$-transformation will be just a complex conjugation, and the Bloch BdG Hamiltonian will be real, so is the magnon wavefunction. For the probe state $W_d^\alpha(\mathbf{q})$, the $PT$-conjugate components are also complex conjugate pair, so in the new basis the probe state $\tilde{W}^\alpha(\mathbf{q})$ is also real. Then we directly apply our discussion about the $Z_2$ charge in electronic band case to the magnon band.

And we know that the $S_z$ conservation will enforce the doubly degeneracy of the magnon band and the nodal lines become Dirac points. We now prove the nodal loop of the WFF on the enclosing sphere will become a nodal point in the presence of the $S_z$ conservation. Before we start, we rearrange the basis in a way

$$\hat{\Psi}(\mathbf{k}) = (\hat{a}_A(\mathbf{k}), \hat{a}_B^\dagger(-\mathbf{k}), \hat{a}_B(\mathbf{k}), \hat{a}_A^\dagger(-\mathbf{k})), \tag{181}$$

so the first $n$ and the last $n$ components have opposite $S_z$ quantum numbers. Here $\hat{a}_{A/B}(\mathbf{k})$ is an $n/2$ dimensional operator vector with $A/B$ denoting the two groups of sublattices related by the inversion operation.

Due to the $U(1)$ symmetry, the Bloch-BdG Hamiltonian in the new basis can be decomposed as

$$H(\mathbf{k}) = \begin{pmatrix} H_+(\mathbf{k}) & \\ & H_-(\mathbf{k}) \end{pmatrix}. \tag{182}$$

The $PT$-symmetry requires that $H_+(\mathbf{k}) = H_-^*(\mathbf{k})$. The doubly degeneracy of the magnon bands can be seen from the fact that if we have

$$H_+(\mathbf{k})\psi_{n\mathbf{k}}^+ = \varepsilon_{n\mathbf{k}}\psi_{n\mathbf{k}}^+, \tag{183}$$

by the $PT$-symmetry, we have

$$H_-(\mathbf{k})\psi_{n\mathbf{k}}^{+*} = (H_+(\mathbf{k})\psi_{n\mathbf{k}}^+)^* = \varepsilon_{n\mathbf{k}}\psi_{n\mathbf{k}}^-, \tag{184}$$

where $\psi_{n\mathbf{k}}^- = \psi_{n\mathbf{k}}^{+*}$. So we have two states with the same energy

$$(\psi_{n\mathbf{k}}, 0), (0, \psi_{n\mathbf{k}}^*), \tag{185}$$

these two states are at different $S_z$ sector and hence, orthogonal. Before we proceed, let us construct two states:

$$\psi_1 = \frac{1}{\sqrt{2}}(\psi_{n\mathbf{k}}, \psi_{n\mathbf{k}}^*); \psi_2 = \frac{i}{\sqrt{2}}(\psi_{n\mathbf{k}}, -\psi_{n\mathbf{k}}^*). \tag{186}$$

Then we transform $H(\mathbf{k})$ to make it explicitly real. Define $U = \frac{1}{\sqrt{2}}\begin{pmatrix} 1 & i \\ 1 & -i \end{pmatrix}$, and we have

$$\tilde{H}(\mathbf{k}) = U^\dagger H(\mathbf{k}) U = \begin{pmatrix} \text{Re} H_+(\mathbf{k}) & -\text{Im} H_+(\mathbf{k}) \\ \text{Im} H_+(\mathbf{k}) & \text{Re} H_+(\mathbf{k}) \end{pmatrix}, \tag{187}$$

and the eigenvectors

$$\tilde{\psi}_1 = (\text{Re}(\psi_{n\mathbf{k}}), \text{Im}(\psi_{n\mathbf{k}})); \tilde{\psi}_2 = (\text{Im}(\psi_{n\mathbf{k}}), -\text{Re}(\psi_{n\mathbf{k}})). \tag{188}$$

Notice that $\tilde{\psi}_2 = i\sigma_2 \tilde{\psi}_1$. Then look at the probe state wavefunction in the new basis:

$$\tilde{W}_1 \propto (\text{Re}(e^{i\mathbf{G}\cdot\mathbf{r}_i}), \text{Im}(e^{i\mathbf{G}\cdot\mathbf{r}_i})); \tilde{W}_2 \propto (\text{Im}(e^{i\mathbf{G}\cdot\mathbf{r}_i}), -\text{Re}(e^{i\mathbf{G}\cdot\mathbf{r}_i})). \tag{189}$$

The final squared overlap matrix is given by

$$M^2 = \begin{pmatrix} A^2 + B^2 & 0 \\ 0 & A^2 + B^2 \end{pmatrix}, \tag{190}$$

where $A = \text{Re}(\psi_{n\mathbf{k}}) \cdot \text{Re}(e^{i\mathbf{G}\cdot\mathbf{r}_i}) + \text{Im}(\psi_{n\mathbf{k}}) \cdot \text{Im}(e^{i\mathbf{G}\cdot\mathbf{r}_i})$, and $B = \text{Re}(\psi_{n\mathbf{k}}) \cdot \text{Im}(e^{i\mathbf{G}\cdot\mathbf{r}_i}) - \text{Im}(\psi_{n\mathbf{k}}) \cdot \text{Re}(e^{i\mathbf{G}\cdot\mathbf{r}_i})$. We can see from here if $\det M^2$ vanishes, the whole $M^2$ will vanish and so there are two constraint conditions, leading to the solutions to forming 0D space (isolated points).

---



\end{document}